\documentclass[journal,onecolumn]{IEEEtran}

\usepackage[font=small]{caption}
\usepackage{cite}
\usepackage{amsmath,amssymb,amsfonts}
\usepackage{amsmath, amsthm}
\usepackage{algorithm}
\usepackage[noend]{algpseudocode}  
\usepackage[bookmarks=false]{hyperref}
\PassOptionsToPackage{bookmarks=false}{hyperref}
\usepackage{pgfplots}
\usepackage{graphicx,enumitem}
\usepackage{textcomp}
\usepackage{xcolor}
\usepackage{mathcomSTEv4}
\usepackage{url}
\usepackage{hyperref}
\usepackage{graphicx}
\usepackage{subcaption}  
\usepackage{multirow}
\usepackage{tikz}

\usetikzlibrary{shapes.geometric, arrows.meta,automata, positioning}
\usepackage{amsmath}
\usepackage{pdflscape}     
\usepackage{booktabs}      

\usepackage{pgfplots}
\usepackage{pgfplotstable}
\pgfplotsset{compat=newest}
\definecolor{forestgreen}{RGB}{3, 9, 184}

\newcommand{\secc}{\textsf{sec}}
\usepackage[labelfont=normalfont,textfont=normalfont]{caption}
\newtheorem{example}{Example}
\AtBeginEnvironment{example}{%
  \pushQED{\qed}%
}
\graphicspath{{images/}}
\AtEndEnvironment{example}{\popQED\endexample}

\newcommand{\algo}{\texttt{PolarZero}}
\newcommand{\Csf}{\mathsf{C}}
\newcommand{\figref}[1]{Fig.~\ref{#1}} 
\newcommand{\tabref}[1]{Table~\ref{#1}}
\newcommand{\secref}[1]{Sec.~\ref{#1}}
\newcommand{\algoref}[1]{Algorithm~\ref{#1}}
\def\BibTeX{{\rm B\kern-.05em{\sc i\kern-.025em b}\kern-.08em
    T\kern-.1667em\lower.7ex\hbox{E}\kern-.125emX}}

\ifCLASSINFOpdf

\else
  
\fi

\usetikzlibrary{arrows.meta, positioning}
\tikzstyle{state}=[circle, draw, minimum size=10pt]

\newcommand{\Sfour}{%
$S_{4} =$
\scalebox{0.8}{$
\left[
\begin{array}{*{4}c}
1&0&0&0\\
1&0&1&0\\
1&1&0&0\\
1&1&1&1\\
\end{array}
\right]
$}
}
\newcommand{\Seight}{%
$S_{8} =$
\scalebox{0.8}{$
\left[
\begin{array}{*{8}c}
1&0&0&0&0&0&0&0\\
1&0&0&0&1&0&0&0\\
1&0&1&0&0&0&0&0\\
1&1&0&0&0&0&0&0\\
1&0&1&0&1&0&1&0\\
1&1&0&0&1&1&0&0\\
1&1&1&1&0&0&0&0\\
1&1&1&1&1&1&1&1\\
\end{array}
\right]
$}
}
\newcommand{\Ssixteen}{%
$S_{16} =$
\scalebox{0.8}{$
\left[
\begin{array}{*{16}c}
1&0&0&0&0&0&0&0&0&0&0&0&0&0&0&0\\
1&0&0&0&0&0&0&0&1&0&0&0&0&0&0&0\\
1&0&0&0&1&0&0&0&0&0&0&0&0&0&0&0\\
1&0&1&0&0&0&0&0&0&0&0&0&0&0&0&0\\
1&1&0&0&0&0&0&0&0&0&0&0&0&0&0&0\\
1&0&0&0&1&0&0&0&1&0&0&0&1&0&0&0\\
1&0&1&0&0&0&0&0&1&0&1&0&0&0&0&0\\
1&1&0&0&0&0&0&0&1&1&0&0&0&0&0&0\\
1&0&1&0&1&0&1&0&0&0&0&0&0&0&0&0\\
1&1&0&0&1&1&0&0&0&0&0&0&0&0&0&0\\
1&1&1&1&0&0&0&0&0&0&0&0&0&0&0&0\\
1&0&1&0&1&0&1&0&1&0&1&0&1&0&1&0\\
1&1&0&0&1&1&0&0&1&1&0&0&1&1&0&0\\
1&1&1&1&0&0&0&0&1&1&1&1&0&0&0&0\\
1&1&1&1&1&1&1&1&0&0&0&0&0&0&0&0\\
1&1&1&1&1&1&1&1&1&1&1&1&1&1&1&1\\ 
\end{array}
\right]
$}
}

\newcommand{\Hsixteen}{%
$H_{16} =$
\scalebox{0.8}{$
\left[
\begin{array}{*{16}c}
1&0&0&0&0&0&0&0&0&0&0&0&0&0&0&0\\ 
1&0&0&0&0&0&0&0&1&0&0&0&0&0&0&0\\ 
1&0&0&0&1&0&0&0&0&0&0&0&0&0&0&0\\ 
1&0&1&0&0&0&0&0&0&0&0&0&0&0&0&0\\  
1&1&0&0&0&0&0&0&0&0&0&0&0&0&0&0\\ 
1&1&0&0&0&0&0&0&1&1&0&0&0&0&0&0\\ 
1&0&1&0&0&0&0&0&1&0&1&0&0&0&0&0\\ 
1&0&0&0&1&0&0&0&1&0&0&0&1&0&0&0\\ 
1&1&1&1&0&0&0&0&0&0&0&0&0&0&0&0\\ 
1&1&0&0&1&0&1&0&0&1&1&0&0&0&0&0\\ 
0&1&1&0&1&1&0&0&1&0&1&0&0&0&0&0\\ 
1&0&1&0&1&0&1&0&1&0&1&0&1&0&1&0\\ 
1&1&0&0&1&1&0&0&1&1&0&0&1&1&0&0\\ 
1&1&1&1&0&0&0&0&1&1&1&1&0&0&0&0\\ 
1&1&1&1&1&1&1&1&0&0&0&0&0&0&0&0\\ 
1&1&1&1&1&1&1&1&1&1&1&1&1&1&1&1\\  
\end{array}
\right]
$}
}

\newcommand{\Anine}{%
$A_9=$
\scalebox{0.8}{%
$
\left[
\begin{array}{*{9}c}
0&0&1&0&0&0&0&0&0\\
0&1&0&0&0&0&0&0&1\\
1&0&1&0&0&0&0&0&0\\
1&1&0&0&0&0&0&0&0\\
0&0&0&0&0&0&0&1&1\\
0&1&0&1&1&0&1&0&0\\
0&0&1&1&0&0&0&1&1\\
1&1&1&1&0&0&1&1&0\\
0&0&1&1&1&1&1&0&1\\
\end{array}
\right]
$}
}
\newcommand{\Aten}{%
$A_{10} =$
\scalebox{0.8}{$
\left[
\begin{array}{*{10}c}
0&0&0&0&0&0&0&0&1&0\\
0&1&0&0&0&0&0&1&0&0\\
0&0&0&0&0&0&1&0&0&1\\
0&0&0&0&0&0&0&1&1&0\\
0&0&0&0&0&0&0&0&1&1\\
1&0&0&0&1&0&1&1&0&0\\
0&0&0&1&1&0&0&0&1&1\\
1&1&0&1&1&0&0&0&0&0\\
0&0&1&1&0&1&1&1&0&1\\
1&1&0&1&1&1&1&0&1&1\\
\end{array}
\right]
$}
}
\newcommand{\Aeleven}{%
$A_{11} =$
\scalebox{0.8}{$
\left[
\begin{array}{*{11}c}
0&0&0&0&0&0&0&0&0&1&0\\
0&0&0&0&1&0&1&0&0&0&0\\
0&0&0&0&0&0&1&0&0&0&1\\
0&0&0&0&0&0&0&0&1&0&1\\
1&1&0&0&0&0&0&0&0&0&0\\
0&0&1&1&0&0&0&1&1&0&0\\
0&0&0&0&0&0&1&1&0&1&1\\
0&0&0&0&0&1&1&0&1&0&1\\
1&0&0&1&0&0&1&1&1&0&1\\
0&1&0&1&0&1&0&1&0&1&1\\
0&1&1&1&1&1&1&0&1&0&1\\
\end{array}
\right]
$}
}
\newcommand{\AtwelveInit}{%
$A_{12}^{[\mathsf{Init}]}[10\!:\!11]$ =
\scalebox{0.8}{$
\left[
\begin{array}{*{12}c}
0&1&1&1&1&0&0&0&1&0&0&0\\
1&1&1&1&1&1&1&1&1&1&1&1\\
\end{array}
\right]
$}
}
\newcommand{\Atwelve}{%
$A_{12} =$
\scalebox{0.8}{$
\left[
\begin{array}{*{12}c}
1&0&0&0&0&0&0&0&0&0&0&0\\
0&0&1&0&0&0&0&1&0&0&0&0\\
1&0&1&0&0&0&0&0&0&0&0&0\\
0&0&1&0&0&1&0&0&0&0&0&0\\
0&0&0&0&0&0&0&0&0&0&1&1\\
1&0&1&0&0&0&0&1&0&1&0&0\\
0&0&0&1&0&1&1&0&0&0&1&0\\
1&0&0&0&0&1&1&0&1&0&0&0\\
0&1&0&0&0&1&1&0&0&1&0&0\\
1&1&1&0&0&1&0&0&0&1&0&1\\
0&1&1&1&1&0&0&0&1&1&0&0\\
1&1&1&1&1&1&1&1&1&1&1&1\\
\end{array}
\right]
$}
}
\newcommand{\Afourteen}{%
$A_{14} =$
\scalebox{0.8}{$
\left[
\begin{array}{*{14}c}
0&0&0&1&0&0&0&0&0&0&0&0&0&0\\
0&0&1&0&0&1&0&0&0&0&0&0&0&0\\
0&0&0&0&1&0&0&1&0&0&0&0&0&0\\
0&0&1&0&0&0&0&0&1&0&0&0&0&0\\
0&0&1&0&0&0&1&0&0&0&0&0&0&0\\
0&0&0&0&0&0&0&1&1&0&0&0&1&1\\
0&0&1&0&1&0&0&0&1&0&0&0&1&0\\
1&1&0&0&0&0&0&1&0&1&0&0&0&0\\
0&0&0&1&1&0&0&1&0&1&0&0&0&0\\
0&0&0&1&1&0&0&1&1&0&1&0&0&1\\
1&1&0&0&1&1&0&0&1&0&0&0&0&1\\
1&1&0&1&1&0&0&1&0&1&0&1&0&1\\
1&1&1&0&0&0&1&1&0&1&1&0&1&0\\
1&0&0&1&0&1&1&1&1&0&0&0&1&1\\
\end{array}
\right]
$}
}

\newcommand{\AsixteenInit}{%
$A_{16}^{[\mathsf{Init}]}[14\!:\!15] =$
\scalebox{0.8}{$
\left[
\begin{array}{*{16}c}
0&0&0&1&1&0&0&0&0&1&1&0&0&0&0&0\\
1&1&1&1&1&1&1&1&1&1&1&1&1&1&1&1\\
\end{array}
\right]
$}
}

\newcommand{\Asixteen}{%
$A_{16} =$
\scalebox{0.8}{$
\left[
\begin{array}{*{16}c}
1&0&0&0&0&0&0&0&0&0&0&0&0&0&0&0\\
0&0&1&0&0&0&0&0&0&0&0&1&0&0&0&0\\
0&0&0&1&0&1&0&0&0&0&0&0&0&0&0&0\\
0&0&0&0&1&0&1&0&0&0&0&0&0&0&0&0\\
0&0&0&0&1&1&0&0&0&0&0&0&0&0&0&0\\
0&1&0&0&1&0&0&0&1&0&0&0&0&0&1&0\\
1&0&0&1&0&1&0&1&0&0&0&0&0&0&0&0\\
1&1&0&0&0&0&1&1&0&0&0&0&0&0&0&0\\
1&1&1&1&0&0&0&0&0&0&0&0&0&0&0&0\\
0&1&0&1&1&1&0&0&1&0&0&0&1&0&0&0\\
0&0&0&0&1&0&0&1&1&0&0&0&1&1&1&0\\
1&1&1&1&1&1&1&1&0&0&0&0&0&0&0&0\\
0&0&1&1&0&0&1&1&1&1&1&1&0&0&0&0\\
1&1&1&1&0&0&0&0&0&0&1&1&0&1&1&0\\
1&0&0&1&1&0&1&0&0&1&1&0&0&1&0&1\\
1&1&1&1&1&1&1&1&1&1&1&1&1&1&1&1\\
\end{array}
\right]
$}
}

\newcommand{\AsixteenHand}{%
$A_{16h} =$
\scalebox{0.8}{$
\left[
\begin{array}{*{16}c}
0&0&0&0&0&1&0&0&0&0&0&0&0&0&0&0\\
0&0&0&1&0&0&0&0&0&0&0&0&1&0&0&0\\
1&0&0&0&1&0&0&0&0&0&0&0&0&0&0&0\\
0&1&1&0&0&0&0&0&0&0&0&0&0&0&0&0\\
0&0&1&1&0&0&0&0&0&0&0&0&0&0&0&0\\
1&0&0&0&1&0&0&0&0&0&0&1&0&0&0&1\\
0&0&0&0&0&1&1&0&0&1&1&0&0&0&0&0\\
0&0&0&0&1&1&0&0&1&1&0&0&0&0&0&0\\
0&0&0&0&1&1&1&1&0&0&0&0&0&0&0&0\\
1&0&0&1&1&0&0&1&1&1&0&0&0&0&0&0\\
0&1&0&1&1&1&0&0&0&1&0&1&0&0&0&0\\
1&0&1&0&0&1&1&0&1&0&0&1&1&0&1&0\\
1&1&0&0&1&1&0&0&1&1&0&0&1&1&0&0\\
1&1&1&1&0&0&0&0&1&1&1&1&0&0&0&0\\
1&1&1&1&1&1&1&1&0&0&0&0&0&0&0&0\\
1&1&1&1&1&1&1&1&1&1&1&1&1&1&1&1\\
\end{array}
\right]
$}
}

\begin{document}

\title{
\algo: A Reinforcement Learning Approach for Low-Complexity Polarization Kernel Design
\thanks{Y. Hong and S. Rini are with the Department of Electrical and Computer Engineering, National Yang Ming Chiao Tung University (NYCU), Hsinchu, Taiwan (emails: \{eltonhong.ee12, stefano.rini\}@nycu.edu.tw). 
L. Barletta is with the Department of Electronics, Information and Bioengineering (DEIB), Politecnico di Milano, Milan, Italy (email: luca.barletta@polimi.it). Part  of  this  work was  presented  at  the 2025 IEEE Global Communications Conference \cite{Rini2512:Reinforcement}. The work of Y.H. and S.R. is supported by the 
NSTC grants number 114-2923-E-A49 -006 -MY3 and 111-2221-E-A49 -068 -MY3.}
}

\author{
Yi-Ting Hong, Stefano Rini, and Luca Barletta
}


\maketitle

\begin{abstract}
%
%
Polar codes with large kernels can achieve improved error exponents but are challenging to design with low decoding complexity. This work investigates kernel construction under recursive maximum likelihood decoding (RMLD) using a reinforcement learning framework based on the Gumbel AlphaZero algorithm. The proposed method efficiently explores the design space and identifies large-size kernels that satisfy a given error exponent while minimizing decoding complexity. For a size-16 kernel, it achieves 17\% lower decoding complexity than handcrafted designs while reaching an error exponent of 0.5183 compared to 0.5 for Arıkan’s kernel, demonstrating the effectiveness of the learning-based approach for practical polar code construction.
\end{abstract}


%
\IEEEpeerreviewmaketitle
\section{Introduction}
%
%
%
%

\IEEEPARstart{P}{olar} codes, introduced by Arıkan~\cite{Ari09}, achieve capacity for any binary-input memoryless symmetric channel. While the original construction employs a \(2 \times 2\) kernel, it is known that larger kernels can improve the error exponent \cite{KorErrExp09}. 
%
However, the joint design of large kernels that simultaneously yield strong polarization properties and low decoding complexity is a challenging problem, primarily due to the exponential growth of the search space.
More generally, design criteria such as computational complexity, memory usage, and parallelism do not lend themselves to tractable mathematical formulations. 
Balancing these aspects is typically a task reserved for experienced domain experts.
In this work, we explore the use of artificial intelligence to automate this design process. 
Our objective is to construct polarization kernels that (i) achieve near-optimal error exponents and (ii) minimize decoding complexity under Recursive Maximum Likelihood Decoding (RMLD) \cite{Lin_RMLD_1998,FujRMLD98,Tri_RTPA_2021,TriRTPA23}.  
%
To this end, we propose an approach based on the Gumbel AlphaZero algorithm, a reinforcement learning (RL) framework that efficiently explores the space of kernel matrices \cite{alphazero,gumble_az}.  
%
%
The proposed method, which we term \algo, discovers kernels that match or surpass handcrafted designs. For size-16 kernels, \algo \space discovers a kernel that achieves a 17\% reduction in RMLD complexity compared to the handcrafted counterpart with the same error exponent.
We argue that \algo~opens new avenues for the data-driven design of polarization kernels, enabling the discovery of codes that meet a variety of performance criteria across multiple dimensions.
More broadly, our results pave the way for the automated design of coding schemes that account for practical implementation constraints such as decoding latency, memory footprint, and parallelism which typically lie beyond the scope of theoretical analysis.

\subsection{Related Work}
In the following, we focus on three main aspects of polar code performance that are relevant to the development of the paper: (i) error exponents (ii) scaling exponent and (iii) decoding complexity.

\noindent
\textbf{(i)} Regarding error exponents, \cite{KorErrExp09} studies large-kernel polar codes and shows that any \( \ell \times \ell \) matrix without an upper triangular column permutation polarizes symmetric channels. The authors derive bounds on the achievable error exponent and show that no kernel of size \( <15 \) exceeds the original \( 2 \times 2 \) kernel's exponent of \( 1/2 \). A BCH-based construction is proposed that asymptotically approaches exponent 1. 
In \cite{pdp_bound}, code decompositions are used to construct nonlinear kernels with improved exponents. They provide optimal constructions for sizes 14–16 by meeting a new upper bound. 

\noindent
\textbf{(ii)} Scaling exponents are another critical factor influencing the performance of polar codes. The finite-length scaling behavior for Arıkan's kernel was studied in \cite{HasScaExp2x214}, where the blocklength required to maintain a target error rate as the information rate approaches capacity was shown to scale as \( (I(W)-R)^{-\mu} \), with \( \mu \approx 3.627 \). In \cite{FazScaExp14} it was shown that improved scaling exponents can be achieved using kernels of size greater than 7. A size-64 kernel with a scaling exponent $\mu\approx2.87$ was reported in \cite{64_yao}.  

\noindent
\textbf{(iii)} From a complexity perspective, RMLD is based on trellis-based ML decoding, originally introduced in \cite{FujRMLD98}, and builds on dynamic programming frameworks such as \cite{ForneyMITnotes}. The first application to polar codes appears in \cite{TriTre19}. 
The recursive trellis processing algorithm (RTPA) in \cite{Tri_RTPA_2021,TriRTPA23} generalizes this approach for large kernels by computing log-likelihood ratios through a recursive trellis structure, achieving lower complexity than full Viterbi decoding. In \cite{TroBrute24}, a depth-first search strategy improves lower bounds on the exponent for kernel sizes 17–29, producing kernels compatible with RTPA. 
Handcrafted kernels optimized for efficient decoding are proposed in \cite{handcraft_win}, which shows that certain \(16 \times 16\) designs outperform Arıkan’s kernel in both exponent and scaling, though at moderately higher complexity. These kernels benefit from structural similarity to  Arıkan's kernel, allowing partial reuse of decoding logic while remaining practical for RTPA.
%
%
In \cite{64_yao}, a size-64 kernel with a scaling exponent of approximately 2.87 was found. However, the decoding complexity of such kernel is very high. In \cite{tri64}, it is shown that the decoding complexity can be reduced by slightly increasing the scaling exponent. The size 64 kernels were constructed using the generalized concatenated code from some size $4\times 16$ or $8\times 8$ inner codes and outer codes. 

\subsection{Contributions}

The design of polarization kernels remains an active area of research, driven by the need for constructions that address practical implementation constraints such as decoding latency and hardware efficiency. This paper takes concrete steps in this direction by focusing on the design of polar codes with large kernels under a decoding complexity constraint.

Previous works have relied either on handcrafted designs or exhaustive search to discover large polarization kernels with desirable properties. In contrast, we propose a reinforcement learning-based method, leveraging the AlphaZero framework, to automatically discover kernels that minimize computational complexity under RMLD.

We implement a tailored version of the RMLD and construct a search environment compatible with AlphaZero.  Our current setup demonstrates that even a basic RL agent can discover a \(16 \times 16\) kernel with lower decoding complexity than that obtained through random search methods.

\begin{figure}[t]
    \centering
    \includegraphics[width=0.95\linewidth]{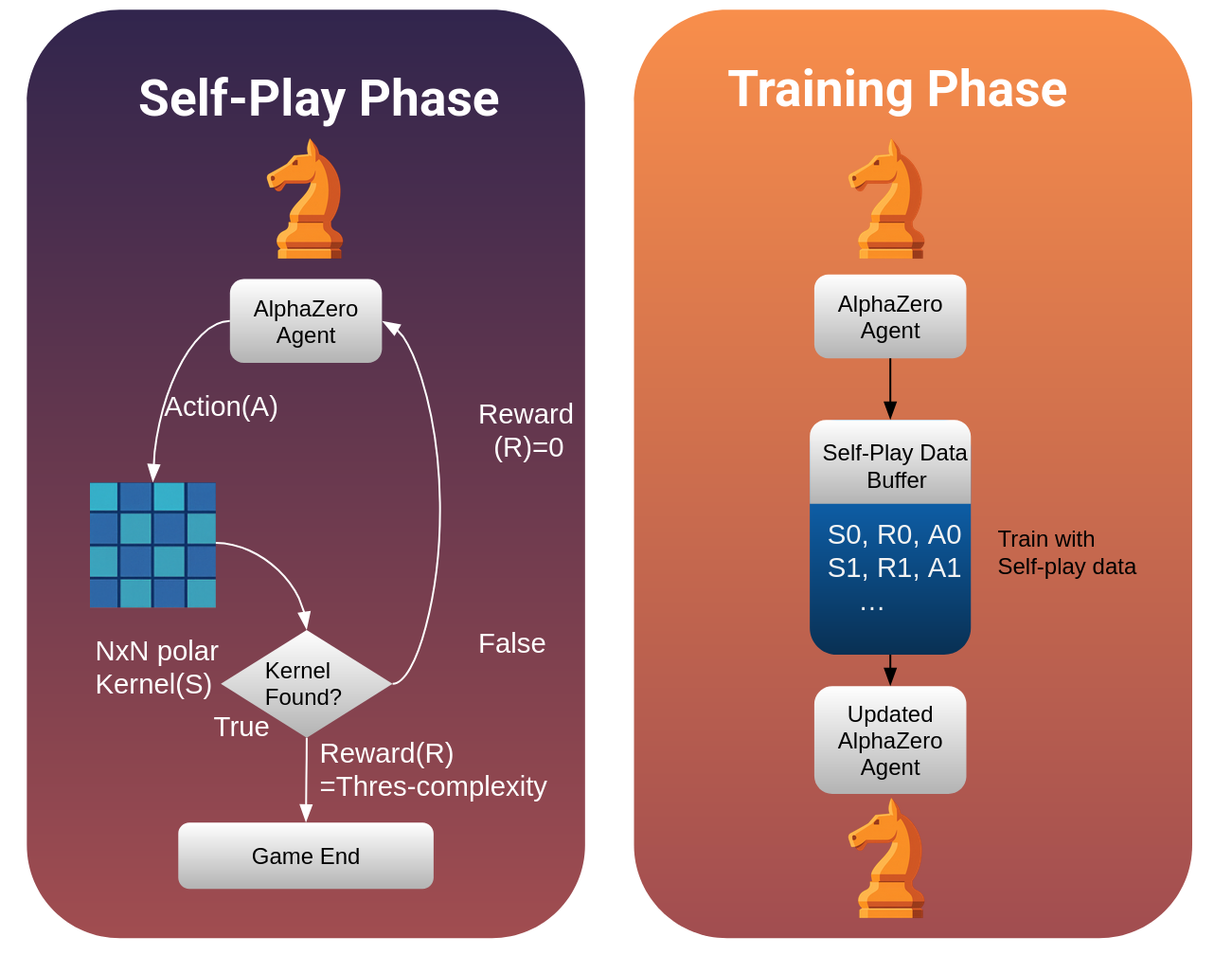}
    
    \caption{A graphical representation of the proposed reinforcement learning algorithm for the design of polarization kernels.In the self-play phase, the agent interacts with the environment and generates self-play data consists of a series of states, actions, rewards. In the training phase, we train the agent using the self-play data obtained from self-play phase to improve the agent policy. }
    \label{fig:graphical_abstract}
\end{figure}

\figref{fig:graphical_abstract} illustrates the architecture of our training framework. During the self-play phase, the RL agent receives the current state of the polar kernel and proposes an action, interpreted as a bit flip in the kernel matrix. The reward is computed as a decreasing function of the kernel’s RMLD complexity. In the training phase, the neural network is updated using data collected through self-play. The objective of the agent is to maximize the cumulative reward, which corresponds to minimizing decoding complexity while satisfying design constraints.

The remainder of the paper is organized as follows:

\begin{itemize}[leftmargin=*,noitemsep]
    \item {\bf Sec.~\ref{sec:Preliminaries}} presents relevant background on polar codes, error exponents, and the RMLD algorithm. A fully worked-out example is included to illustrate the complexity calculation process and to provide intuition for the proposed design criteria.

    \item {\bf Sec.~\ref{sec:Problem Formulation}} formally defines the low-complexity kernel design problem as a constrained optimization task over binary $\ell \times \ell$ matrices that satisfy a target error exponent.

    \item {\bf Sec.~\ref{sec:Proposed Approach}} introduces our reinforcement learning-based method, \algo, including details of the kernel construction environment, the AlphaZero framework, and reward shaping techniques tailored to the decoding complexity metric. Also, two algorithmic extensions are explored. In the first, we initialize the bottom rows of the kernel with a fixed, handcrafted submatrix to guide the learning process. In the second, we train kernels of multiple sizes within a single model.

\end{itemize}
Finally, Sec. \ref{sec:conclusion} concludes the paper.

Through these components, we demonstrate that \algo~offers a scalable, flexible, and data-driven framework for the design of polarization kernels. \algo~navigates the high-dimensional design space beyond the reach of brute-force or structural heuristics, enabling the automated discovery of practical polar codes that are both efficient to decode and strong in error performance.

\subsection{Notation}
Calligraphic letters denote sets. The set \( \{0, 1, \dots, n-1\} \) is denoted by \( [n] \), and \( \{n, \dots, m\} \triangleq [n,m] \)  for \( n < m \), and $\{n, \dots, m-1\} \triangleq [n,m)$.
%
For a vector \(x\), we denote by \( x_k^{n-1} = (x_k, \ldots, x_{n-1}) \) 
the sub-vector consisting of elements from index \(k\) to \(n-1\).
%
For a vector $x$, both $x_i$ and $x[i]$ denote its $i$-th element. Similarly, for a matrix $y$, both $y_i$ and $y[i]$ denote its $i$-th row. 
Moreover, both \( y_i^{k} \) and \( y[i:k] \)  denote the submatrix
\[
y[i:k] =
\begin{bmatrix}
y_i \\
y_{i+1} \\
\vdots \\
y_k
\end{bmatrix}.
\]

Random variables are denoted by uppercase letters, and their realizations by lowercase.
%
%
The all-zero matrix of size \( p \times q \) is denoted \( 0^{p \times q} \). The all-zero vector of size \(p\) is denoted \(0^p\). 
$[\av,\bv]$ denotes the horizontal concatenation of the vectors $\av$ and $\bv$.
Similarly, let $M$ be a matrix of size $p \times q_1$ and $N$ a matrix of size $p \times q_2$. Then $[M, N]$ denotes the horizontal concatenation of $M$ and $N$, yielding a matrix of size $p \times (q_1+q_2)$.
The Kronecker product between two matrices is denoted by $\otimes$, and the $m$-fold Kronecker product of a matrix $A$ by $A^{\otimes m}$.
$\Pr(\mathcal{X})$ denotes the probability of an event $\mathcal{X}$.
 \section{Preliminaries}
\label{sec:Preliminaries}

 \subsection{Polar Codes}
 \label{sec:Polar Codes}

A polar code over \( \mathbb{F}_q \) is defined by the transformation \( c_0^{n-1} = \hat{u}_0^{n-1} K^{\otimes m} \), where \( K \) is a non-singular \( \ell \times \ell \) kernel matrix, and \( n = \ell^m \). 
The entries of vector \( \hat{u}_0^{n-1} \in \mathbb{F}_q^n \) indexed by the frozen set \( \mathcal{F} \subset [n] \) satisfy \( \hat{u}_i = 0 \) for all \( i \in \mathcal{F} \), while the remaining entries carry information symbols. 
This definition can be generalized to obtain mixed kernel polar codes with the codewords given by 
\[
c_0^{n-1} = \hat{u}_0^{n-1} (K_{l_1} \otimes K_{l_2} \otimes \cdots \otimes K_{l_m}),
\]
where \( K_{l_i} \) is a kernel of dimension \( l_i \) \cite{presman2015mixed}. 
The multi-kernel approach described in \cite{MultiKern} incorporates kernels of different sizes (including length-3 and length-5 kernels) that are decoded through specialized f-function and g-function operations during successive-cancellation (SC) decoding.
Polarization conditions for binary kernels were established in~\cite{KorErrExp09} and extended to the non-binary case in~\cite{mori2014source}. 
Throughout this work, we focus exclusively on binary polar codes (i.e., \( q = 2 \)) and assume that identical kernels are used across all levels of the polar transformation.

\subsection{Error Exponent for Large Kernels}
\label{sec:Error exponent for large kernel}

Two common criteria for evaluating the performance of a polarization kernel \( K \) are the \emph{scaling exponent}~\cite{HasScaExp2x214, FazScaExp14} and the \emph{error exponent}~\cite{KorErrExp09}. In this work, we focus on the error exponent as the primary performance metric.

Let \( W: \{0,1\} \rightarrow \mathcal{Y} \) be a symmetric binary-input discrete memoryless channel (B-DMC) with capacity \( I(W) \), and let \( Z_m \) denote the Bhattacharyya parameter of a subchannel chosen uniformly at random among those induced by the polar transform \( K^{\otimes m} \). Then, the kernel \( K \) is said to have error exponent \( E(K) \) if:
\begin{itemize}
    \item[(i)] For any fixed \( \beta < E(K) \), \( \liminf_{m \to \infty} \Pr\left[ Z_m \leq 2^{-\ell^{m \beta}} \right] = I(W) \).
    \item[(ii)] For any fixed \( \beta > E(K) \), \( \liminf_{m \to \infty} \Pr\left[ Z_m \geq 2^{-\ell^{m \beta}} \right] = 1 \).
\end{itemize}
As a consequence, if the frozen set \( \mathcal{F} \) selects the \( n(1-R) \)  subchannels with worst Bhattacharyya parameter, where $R \in (0,I(W))$ is the code rate, then for any fixed \( \beta < E(K) \) the block error probability under successive cancellation decoding satisfies:
\[
P_e(\mathcal{F}, W) \leq 2^{-n^\beta}
\]
for sufficiently large \( n \).
   
The authors of \cite{KorErrExp09} show that the error exponent $E(K)$ can be expressed as a function of the \emph{partial distance profile} (PDP)   of $K$. The PDP is denoted as \( \Dv(K) = (D_0, \dots, D_{\ell-1}) \), where
\begin{equation}\label{eq:def_Di}
D_i = \min_{\mathbf{c} \in \langle K_{i+1}^{\ell-1} \rangle} d_H(K_i, \mathbf{c}), \quad 0 \leq i < \ell,   
\end{equation}
with \( K_i \) being the \( i \)-th row of \( K \), \( d_H \) the Hamming distance and $\mathbf{c}$ the codewords generated from matrix $K_{i+1}^{\ell-1}$. Specifically, \cite{KorErrExp09} shows that the error exponent associated with kernel $K$, denoted by \( E(K) \), is given by:
\ea{
E(K) = \frac{1}{\ell} \sum_{i=0}^{\ell-1} \log_\ell D_i.
\label{eq:E_K}
}
When needed, we will explicitly indicate the dependence of the error exponent from the PDP with the slight abuse of notation as $E(\Dv)$ instead of $E(K)$ as in \eqref{eq:E_K}. 
Note that $\widetilde{\Dv}(\ell)$ denotes the target PDP of size $\ell$, as specified in Table~\ref{tab:pdp}, whereas $\Dv(K)$ represents the PDP  obtained from the kernel $K$.

%
%

\subsection{Partial Distance Profile (PDP)}
\label{sec:PDP}

In~\cite{pdp_bound}, a linear programming (LP) approach is proposed to upper bound the maximum achievable error exponent for a kernel of size \( \ell \). This bound is expressed via a bounding PDP sequence \( \bar{\Dv}(\ell) = (\bar{D}_0, \dots, \bar{D}_{\ell-1}) \). 

While every achievable PDP \( \Dv(K) \) satisfies \( \Dv(K) \leq \bar{\Dv}(\ell) \) (component-wise), the converse does not necessarily hold.
%
Therefore, practical kernel search methods typically use relaxed profiles \( \widetilde{\Dv}(\ell) \leq \bar{\Dv}(\ell) \) as design targets. The relaxed profiles used in our experiments for kernel sizes \( \ell \in [5,16] \) are listed in Table~\ref{tab:pdp}. 

\begin{table}[tbp]
\caption{\centering Relaxed PDP \( \widetilde{\Dv}(\ell) \) and corresponding error exponents. \\
Here, we indicate the error exponent as $E(\Dv)$ instead of $E(K)$ as in \eqref{eq:E_K}. 
}

\begin{center}
\begin{tabular}{|c|l|c|c|}
\hline
\( \ell \) & \rule{0pt}{2.3ex} \( \widetilde{\Dv}(\ell) \) & \( E(\widetilde{\Dv}) \) & \( E(\bar{\Dv}) \) \\
\hline
5  & 1,2,2,2,4 & 0.4307 & 0.4307 \\
6  & 1,2,2,2,4,4 & 0.4513 & 0.4513 \\
7  & 1,2,2,2,4,4,4 & 0.4580 & 0.4580 \\
8  & 1,2,2,2,4,4,4,8 & 0.5000 & 0.5000 \\
9  & 1,2,2,2,2,4,4,6,6 & 0.4616 & 0.4616 \\
10 & 1,2,2,2,2,4,4,4,6,8 & 0.4692 & 0.4692 \\
11 & 1,2,2,2,2,4,4,4,6,6,8 & 0.4775 & 0.4775 \\
12 & 1,2,2,2,2,4,4,4,4,6,6,12 & 0.4825 & 0.4961 \\
13 & 1,2,2,2,2,4,4,4,4,6,6,8,10 & 0.4883 & 0.5005 \\
14 & 1,2,2,2,2,4,4,4,4,6,6,8,8,8 & 0.4910 & 0.5019 \\
15 & 1,2,2,2,2,4,4,4,4,6,6,8,8,8,8 & 0.4978 & 0.5077 \\
16 & 1,2,2,2,2,4,4,4,4,6,6,8,8,8,8,16 & 0.5183 & 0.5274 \\
\hline
\end{tabular}
\label{tab:pdp}
\end{center}
\end{table}


The simplest approach for computing the partial distance \( D_i \) for a kernel \( K \in \mathbb{F}_2^{\ell \times \ell} \) is by full enumeration: according to~\eqref{eq:def_Di}, generate all codewords in the subcode \( \langle K_{i+1}^{\ell-1} \rangle \), sum them with \( K_i \), and return the minimum Hamming weight among the resulting vectors.
However, the number of codewords in \( \langle K_{i+1}^{\ell-1} \rangle \) grows exponentially with \( \ell \), making this method impractical for \( \ell > 32 \).

An alternative method relies on RMLD-based list decoding. To compute \( D_i \), one transmits \( K_i \) as a codeword over a noiseless channel and decodes it using the RMLD list decoder (with list size \( L \)) over the subcode generated by \( K_{i+1}^{\ell-1} \). The output list \( {\cal L} \) is then used to compute an estimate of $D_i$ as 
\( \hat{D}_i = \min_{\mathbf{c} \in {\cal L}} d_H(K_i, \mathbf{c}) \). 
If the list size $L$ is sufficiently large, then \( \hat{D}_i = D_i \).

While list decoding also grows in complexity, it provides a tractable alternative for moderate kernel sizes (e.g., \( \ell \in [32, 64] \)). In this paper, since we focus on \( \ell \leq 16 \), we use full enumeration via a depth-first search (DFS), as in~\cite{TroBrute24}, to compute PDP values.

\section{Recursive Maximum Likelihood Decoding (RMLD)}
\label{sec:Recursive Maximum Likelihood Decoding}

Recursive Maximum Likelihood Decoding (RMLD), originally proposed in~\cite{FujRMLD98}, is a trellis-based algorithm for decoding binary linear block codes that achieves maximum likelihood (ML) performance with significantly reduced complexity. 
Unlike conventional Viterbi decoding, which constructs the full code trellis, RMLD recursively builds compact metric tables based on small, one-section trellises with limited state and branch complexity. The recursion proceeds via two main steps: a base case initialization and a recursive update. This structure allows RMLD to preserve ML optimality while substantially reducing computational cost.
The RMLD framework has been extended to polar codes with large kernels through the Recursive Trellis Processing Algorithm (RTPA)~\cite{Tri_RTPA_2021,TriRTPA23}. 
RTPA leverages the recursive structure of polar codes, which can be interpreted as generalized concatenated codes~\cite{blokh1974coding} composed of non-systematic inner codes. This interpretation enables efficient decoding by reusing substructures in the kernel and sharing computation across decoding phases.

Due to the central role of RMLD in the remainder of this work, we next provide a high-level overview of its core ideas. For further background on trellis-based decoding, including RMLD, we refer the reader to~\cite[Ch.~6 and~11]{Lin_RMLD_1998}.

\subsection{High-Level Perspective}
The key idea of RMLD is to exploit the symmetries of the minimal trellis associated with a linear block code to reduce decoding complexity.
Given any linear block code, one can construct a minimal trellis for ML decoding. However, this trellis can often be ``sectionalized'' so that transitions between states correspond to multiple code bits. In this form, a single branch may represent several code bits, and two adjacent states may be connected by multiple branches.

Each group of such branches corresponds to a coset of a subcode. The RMLD algorithm processes these \emph{composite branches} by constructing a \emph{composite branch metric table}, which stores, for each coset: (i) the highest path metric (computed recursively), and (ii) the label of the codeword achieving this metric.
By combining the metric tables for shorter trellis sections, RMLD recursively builds the table for a longer section. This divide-and-conquer strategy drastically reduces the number of paths that need to be considered compared to Viterbi decoding.
In essence, RMLD performs optimal ML decoding by processing only the distinct coset representatives in each trellis section, thereby achieving complexity reduction through structural reuse.

\subsection{RMLD for Polar Codes}

In~\cite{TriRTPA23}, the RMLD algorithm is adapted to the decoding of polar codes with large kernels. While RMLD achieves significantly lower complexity than Viterbi decoding, its naive implementation still suffers from exponential growth in complexity with respect to the number of cosets in the kernel. This complexity remains far higher than the \( O(\ell \log \ell) \) complexity of Arıkan's successive cancellation (SC) decoding.
To address this,~\cite{TriRTPA23} introduces the Recursive Trellis Processing Algorithm (RTPA), a refinement of RMLD tailored to the recursive structure of polar codes. RTPA reduces complexity via two key strategies:
\begin{itemize}[leftmargin=*,noitemsep]
    \item {Trellis reuse:} intermediate trellis sections are reused across different decoding phases, avoiding redundant computation;
    \item {Specialized LLR-domain processing:} for certain structured kernels, closed-form expressions can be derived for composite branch metrics, allowing them to be computed with simple comparisons and additions instead of full metric aggregation.
\end{itemize}
The second technique exploits structural patterns in the kernel matrix to simplify computation. In particular,~\cite[Sec.~III-C]{TriRTPA23} identifies several \emph{special trellis cases} in which the recursive updates in the LLR domain can be implemented using only 1 or 2 comparisons and a small number of additions. For example, when the generator rows involved satisfy certain symmetry or orthogonality conditions, all entries in a composite branch metric table (CBT) can be shifted by a constant without affecting the decoding decision. This enables a simplified CBT computation using only relative metric differences. 

Although RTPA still exhibits exponential complexity in the worst case, several kernels admit efficient trellis reuse and special LLR-domain processing that bring their decoding complexity close to that of SC decoding. For instance, the sorted Arıkan kernel achieves the same complexity under RTPA as in SC decoding, while maintaining the same error exponent (\( E = 0.5 \))~\cite[Sec.~IV]{Tri_RTPA_2021,TriRTPA23}. Furthermore, some handcrafted kernels improve the error exponent beyond \( 0.5 \) with only a moderate increase in complexity~\cite{handcraft_win,tri64}.

RTPA thus enables a flexible trade-off between decoding complexity and error performance, opening the door to practical use of large kernels in polar code design.







\subsection{Notation and Preliminaries}
\label{sec:notation}

Before proceeding with the decoding strategies for specific kernels, we introduce the main notation used throughout the remainder of the section.
%
%
Define  $\secc(x\!:\!y)$ as section in position $[x,y)$. 
Consider a binary $(n,k)$ linear block code $\mathcal{C}$. 
%
%
For two nonnegative integers \( x \) and \( y \) such that \( 0 \le x \le y \le n \), let \( \mathcal{C}_{xy} \) denote the subcode of \( \mathcal{C} \) whose codewords are zero everywhere except in the section \( \secc(x\!:\!y) \).
%
\( p_{xy}(\mathcal{C}) \) denotes the punctured code of \( \mathcal{C} \) obtained by deleting all coordinates outside \( \secc(x\!:\!y) \).
%
%
$s_{xy}(\mathcal{C}) = p_{xy}(\mathcal{C}_{xy})$ denotes the shortened code of \( \mathcal{C} \) obtained by puncturing all coordinates outside \( \secc(x\!:\!y) \) from the codewords in \( \mathcal{C}_{xy} \).
Therefore, $s_{xy}(\mathcal{C})$ is a linear subcode of $p_{xy}(\mathcal{C})$.   For a linear code $A$ and a linear subcode $B$ of $A$, $A/B$ denotes the set of cosets of $B$ in $A$. The list of notations can be found in Table~\ref{tab:sc_notation}.    


For a matrix $A \in \mathbb{R}^{m \times n}$, we use the notation
\[
A \leftarrow \! \! \! \lcb \p{r_1 \\ r_2 \\ \vdots \\ r_m} \rnone
\]
to indicate that the rows of $A$, from top to bottom, are indexed by
$r_1, r_2, \ldots, r_m$. The Hamming weight of a binary vector $a$ is defined as $w_H(a)$, and the Hamming distance between two binary vectors $a$ and $b$ is defined as $d_H(a,b)$.

The trellis-oriented generator matrix (i.e., a minimum-span generator matrix) is denoted by $\mathbb{G}$ \cite{KscTreG95}. For a row $G_i$ in a generator matrix $G$, we define its start position as the index of the first nonzero element and its end position as the index of the last nonzero element. In a trellis-oriented generator matrix $\mathbb{G}$, no two rows share the same start position, and no two rows share the same end position. An arbitrary generator matrix $G$ can be transformed into a trellis-oriented generator matrix $\mathbb{G}$ using Gaussian elimination, as described in \cite[Sec.~IV-C]{KscTreG95}.

When applied to polar codes, a size-\( \ell \) kernel \( G_\ell \) induces \( \ell \) decoding phases under the RMLD framework. In each phase \( i \in [\ell] \), decoding is performed using an \emph{extended kernel} \( G^{(i)}_\ell \) constructed as follows:
\begin{enumerate}
    \item Remove the first \( i \) rows of \( G_\ell \), and
    \item Append a column \( \kappa \in \mathbb{F}_2^{\ell - i} \) with a leading 1 followed by zeros:
    \ea{
    \kappa = \lsb 1 , 0^{\ell-i-1} \rsb^T.
    \label{eq:kappa}
    }
\end{enumerate}
The resulting extended kernel \( G^{(i)}_\ell \) is an \( (\ell - i) \times (\ell + 1) \) matrix.
This construction mirrors the two-section trellis method in~\cite[Sec.~11.4]{Lin_RMLD_1998}, where an appended identity column identifies the coset in the current trellis section.
Let \( \bar{c}^{(i)} \in \mathbb{F}_2^\ell \) denote a codeword generated by the extended generator matrix \( G_\ell^{(i)} \), corresponding to the decoding of bit \( u_i \). Specifically, we define:
\begin{equation}
\bar{c}^{(i)} = u_i^{\ell-1} G_\ell^{(i)}.
\label{eq:extended kernel}
\end{equation}
As \( G_\ell^{(i)} \) has a leading 1 followed by zeros in its last column, it follows that \( \bar{c}^{(i)}_\ell = u_i \).

We adopt the following standard log-likelihood ratio (LLR) notation: 
\begin{subequations}\label{eq:llrs}
\begin{align}
L_j &= \log \frac{\Pr(c_j = 0 \mid y_j)}{\Pr(c_j = 1 \mid y_j)} \\
l_j &= \exp(L_j) = \frac{\Pr(c_j = 0 \mid y_j)}{\Pr(c_j = 1 \mid y_j)} \\
\hat{L}_j &= \log \frac{\Pr(u_j = 0 \mid y_0^{\ell-1}, u_0^{j-1})}{\Pr(u_j = 1 \mid y_0^{\ell-1}, u_0^{j-1})} \\
\hat{l}_j &= \exp(\hat{L}_j) = \frac{\Pr(u_j = 0 \mid y_0^{\ell-1}, u_0^{j-1})}{\Pr(u_j = 1 \mid y_0^{\ell-1}, u_0^{j-1})}.
\end{align}
\end{subequations}
Here, \( y_j \) is the channel output for bit \( c_j \). \( L_j \) and \( \hat{L}_j \) denote the channel and posterior LLRs respectively, and \( l_j, \hat{l}_j \) are their exponential (likelihood ratio) forms.

After obtaining the soft output $\hat{L}_i$, a hard decision is made to determine $\hat{u}_i$, i.e.,
\begin{equation}\label{eq:hard_decision}
\hat{u}_i =
\begin{cases}
0, & \text{if } \hat{L}_i > 0, \\[6pt]
1, & \text{otherwise}.
\end{cases}
\end{equation}

\medskip

In general, the goal of SC-based decoding is to estimate each bit \( u_i \) sequentially via maximum likelihood:
\begin{equation}
\hat{u}_i = \argmax_{u_i \in \{0,1\}} \sum_{\bar{c}^{(i)} \in \{u_i G_\ell[i]+ \langle G_\ell^{(i+1)}\rangle\}} \prod_{j=0}^{\ell-1} \Pr(c_j = \bar{c}^{(i)}_j \mid y_j),
\label{eq:sc_decoding0}
\end{equation}
where $G^{(\ell)}_\ell$ is understood as an empty matrix.
Note that the summation is taken over the coset \( u_i^\ell G_\ell^{(i)} \), i.e., the set of codewords consistent with the fixed value of \( u_i \). This structure enables a recursive formulation for efficient SC decoding.


\subsection{Successive Cancellation (SC) Decoding of Arıkan's Kernel}

To illustrate the decoding process, we begin by reviewing SC decoding for the classical Arıkan kernel of dimension \( \ell = 2 \).
%
%
Consider the \( 2 \times 2 \) Arıkan kernel:
\begin{equation}
F_2 =
\begin{bmatrix}
1 & 0 \\
1 & 1
\end{bmatrix},
\label{eq:F2}
\end{equation}
which achieves channel polarization under successive cancellation (SC) decoding.

Under SC decoding with Arıkan’s kernel, the decoding process proceeds sequentially as follows:
\begin{itemize}
    \item Phase \( i = 0 \): Apply the \emph{F-function}:
    \begin{equation}\label{f_func_approx}
    \hat{L}_0 = \log\left( \frac{e^{L_0 + L_1} + 1}{e^{L_0} + e^{L_1}} \right) 
    \approx \text{sign}(L_0)\text{sign}(L_1)\min\{|L_0|, |L_1|\}.
    \end{equation}

    If \( u_0 \) is a frozen bit or \( \hat{L}_0 > 0 \), set \( \hat{u}_0 = 0 \); otherwise, set \( \hat{u}_0 = 1 \).
    
    \item Phase \( i = 1 \): Apply the \emph{G-function}:
    \begin{equation}
    \hat{L}_1 = L_1 + (-1)^{\hat{u}_0} L_0.
    \end{equation}
    If \( u_1 \) is a frozen bit or \( \hat{L}_1 > 0 \), set \( \hat{u}_1 = 0 \); otherwise, set \( \hat{u}_1 = 1 \).
\end{itemize}

\begin{figure}[!ht]
    \centering
    \begin{tikzpicture}[node distance=1cm, every node/.style={scale=1}]
        \node (L0) [left=0.5cm] {\(L_0\)};
        \node (L1) [below of=L0] {\(L_1\)};
        \node (box) [draw, minimum width=2cm, minimum height=2cm, right of=L0, xshift=1cm, yshift=-0.5cm] {$F_2$};
        \node (L0out) [right of=box, xshift=1cm, yshift=0.5cm] {\(\hat{L}_0\)};
        \node (L1out) [right of=box, xshift=1cm, yshift=-0.5cm] {\(\hat{L}_1\)};
        \node (u0) [right of=L0out, xshift=0.25cm] {\(\hat{u}_0 \)};
        \node (u1) [right of=L1out, xshift=0.25cm] {\(\hat{u}_1 \)};
        \draw[->] (L0) -- (box.west|-L0);
        \draw[->] (L1) -- (box.west|-L1);
        \draw[->] (box.east|-L0) -- (L0out);
        \draw[->] (box.east|-L1) -- (L1out);
        \draw[->] (L0out) -- (u0);
        \draw[->] (L1out) -- (u1);
    \end{tikzpicture}
    \caption{SC decoding of Arıkan’s kernel \( F_2 \): LLR transformation and hard decisions.}
    \label{fig:F2}
\end{figure}
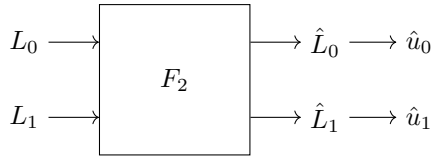

This process is illustrated in \figref{fig:F2}, where the input LLRs are processed through the \( F_2 \) kernel to produce the output LLRs \( \hat{L}_0 \) and \( \hat{L}_1 \), along with their corresponding hard decisions \( \hat{u}_0 \) and \( \hat{u}_1 \).

\subsection{Extended Kernel Construction and Trellis Decoding}
\label{sec: Extended Kernel Construction}

%
We now reinterpret SC decoding through these trellis-based lens, bridging it with the RMLD formulation. This view highlights how the classical \( F \)- and \( G \)-functions emerge naturally from ML decoding over codeword ensembles shaped by the kernel structure.

Let us first define the extended kernels for Arıkan’s \( 2 \times 2 \) kernel: 
\[
F_2 = 
\begin{bmatrix}
1 & 0 \\
1 & 1
\end{bmatrix}.
\label{eq:F2_}
\]

\paragraph{Phase \( i = 0 \)}  
The extended kernel is:
\ea{
F_2^{(0)} =
\lsb
\begin{array}{cc|c}
1 & 0 & 1 \\
1 & 1 & 0
\end{array}
\rsb.
\label{eq:F2_0}
}
The codewords generated by the exteded kernel, $\bar{c}^{(0)} = u_0^1 F_2^{(0)}$, are:
\[
\begin{cases}
[00|0], [11|0] \ \ \ &  \ \ \ \text{for } u_0 = 0, \\
[10|1], [01|1] \ \ \ &  \ \ \ \text{for } u_0 = 1.
\end{cases}
\]

\paragraph{Phase \( i = 1 \)}  
The extended kernel is:
\ea{
F_2^{(1)} =
\left[
\begin{array}{cc|c}
1 & 1 & 1
\end{array}
\right].
\label{eq:F2_1}
}
The corresponding codewords in the exteded kernel, that is $\bar{c}^{(1)} = u_1^1 F_2^{(1)}$,  are:
\[
\begin{cases}
[00|0] & \text{for } u_1 = 0 \\
[11|1] & \text{for } u_1 = 1.
\end{cases}
\]

\bigskip

In both phase $0$ and phase $1$, SC decoding reduces to selecting the most likely bit \( u_i \) given all prior decisions, by solving:
\begin{equation}
\hat{u}_i = \argmax_{u_i \in \{0,1\}} \sum_{\bar{c}^{(i)} \in \{u_i F_2[i]+  \langle F_2^{(i+1)}\rangle\} } \prod_{j=0}^{\ell-1} \Pr(c_j = \bar{c}^{(i)}_j \mid y_j).
\label{eq:sc_decoding1}
\end{equation}

This formulation can be generalized to $\ell$-size kernel \( G_\ell \) by replacing \( F_2^{(i)} \) with the corresponding \( G_\ell^{(i)} \). Specifically,
\begin{equation}
\hat{u}_i = \argmax_{\bar{c}^{(i)}_\ell \in \{0,1\}} \sum_{\bar{c}^{(i)} \in  \{\bar{c}_\ell^{(i)} G_{\ell}[i]+\langle G_\ell^{(i+1)} \rangle\}} \prod_{j=0}^{\ell-1} \Pr(c_j = \bar{c}^{(i)}_j \mid y_j).
\label{eq:sc_decoding}
\end{equation}
%
In \eqref{eq:sc_decoding}, the constraint \( \bar{c}^{(i)}_\ell = u_i \) ensures that the last bit of each candidate codeword matches the hypothesis \( u_i \in \{0,1\} \). This reflects the fact that, due to the appended identity column \( \kappa \) in \eqref{eq:kappa}, the final symbol in any codeword from \( G_\ell^{(i)} \) directly encodes the bit being estimated in phase \( i \).

The argument of the summation  in \eqref{eq:sc_decoding} contains the expression \( \bar{c}^{(i)} \in \{ \bar{c}^{(i)}_\ell G_{\ell}[i] + \langle G_\ell^{(i+1)} \rangle \} \).
%
The first term,
$\bar{c}^{(i)}_\ell G_{\ell}[i] = u_i G_{\ell}[i]$ 
is a vector equal to \( G_\ell[i] \) when \( u_i = 1 \), and equal to \( 0^\ell \) when \( u_i = 0 \). 
The second term, \( \langle G_\ell^{(i+1)} \rangle \), denotes the subcode generated by \( G_\ell^{(i+1)} \). 
The sum of these two terms forms a coset in \( G_\ell^{(i)} / G_\ell^{(i+1)} \). Specifically,
$\langle G_\ell^{(i+1)} \rangle$ when  $u_i = \bar{c}_\ell^{(i)} = 0$ 
and
$G_\ell[i] + \langle G_\ell^{(i+1)} \rangle$ when  $u_i = \bar{c}_\ell^{(i)} = 1$.

Using the extended kernel, we can transform the bit-wise SC decoding decision into a standard trellis decoding problem. Each hypothesis \( u_i = 0 \) or \( u_i = 1 \) corresponds to a coset of codewords in \( \langle G_\ell^{(i)} \rangle \), and the decoder selects the coset whose most likely codeword best matches the received observations. 

In the LLR domain, the MAP expression in~\eqref{eq:sc_decoding} can be approximated as in~\cite{TriRTPA23}. In particular, by dividing the case 
\( u_i = \bar{c}_\ell^{(i)} = 0 \) in~\eqref{eq:sc_decoding} by the case 
\( u_i = \bar{c}_\ell^{(i)} = 1 \), one obtains the likelihood ratio \( \hat{\ell}_i \). 
The corresponding log-likelihood ratio \( \hat{L}_i \) is then approximated using the min-sum approximation.
The resulting expression is given below, and its detailed derivation is provided in Appendix~\ref{app:MAP_der}.
\begin{equation}
\hat{L}_i \approx \max_{\bar{c}^{(i)} \in  \langle G_\ell^{(i) }\rangle \,:\, \bar{c}^{(i)}_\ell = 0}
\sum_{j=0}^{\ell-1} F(L_j \mid \bar{c}^{(i)}_j )
-
\max_{\bar{c}^{(i)} \in \langle G_\ell^{(i)}\rangle \,:\, \bar{c}^{(i)}_\ell = 1}
\sum_{j=0}^{\ell-1} F(L_j \mid \bar{c}_j^{(i)})
\label{eq:llr_sc}
\end{equation} 
where the function \( F(L_j \mid \bar{c}_j^{(i)}) \) selects the contribution of the \( j \)-th LLR depending on the codeword bit \( \bar{c}_j^{(i)} \):

\begin{equation}
F(L_j \mid \bar{c}_j^{(i)}) =
\begin{cases}
L_j, & \text{if } \bar{c}_j^{(i)} = 0, \\
0,   & \text{if } \bar{c}_j^{(i)} = 1.
\label{eq:sc_llr_path_metric}
\end{cases}
\end{equation}
This formulation allows us to approximate MAP decoding over extended generator matrices using additive metrics in the LLR domain.
%
At each phase \( i \), ML decoding is performed with respect to the coset \( G_\ell^{(i)} / G_\ell^{(i+1)} \) in order to generate \( \hat{L}_i \) as in~\eqref{eq:llr_sc}. The structure of \( G_\ell^{(i)} \) guarantees that the bit \( u_i \) can be read directly from the final symbol of \( \bar{c}^{(i)} \),  enabling implementation using trellis-based ML decoding techniques such as the Viterbi algorithm or RMLD. 
%
In particular, \eqref{eq:llr_sc} can be solved using the Viterbi algorithm (dynamic programming), where each path accumulates its metric according to~\eqref{eq:sc_llr_path_metric}, and a maximization is performed to select the path converging to the same node.
In Appendix \ref{app:size 4} we present an example of the extended kernel construction for a kernel of size 4.

\subsection{RMLD for Larger Kernels}\label{Sec:RMLD_steps_}

Although the Viterbi algorithm can be used to solve \eqref{eq:llr_sc}, its computational complexity grows exponentially with the kernel size and quickly becomes impractical for larger kernels. In contrast, RMLD achieves lower complexity by partitioning the trellis into multiple sections and by applying a divide-and-conquer approach. This method recursively decomposes the generator matrix into two parts, enabling more efficient decoding. 
This is comprised of the following steps -- each detailed in the corresponding section below:
\begin{itemize}
    \item 
    \emph{\underline{Step 1:}} Punctured Code Construction
    \item 
    \emph{\underline{Step 2}:} Mapping Table Construction
    \item 
    \emph{\underline{Step 3:}} Max Tree Construction
    \item
    \emph{\underline{Step 4:}} RMLD Decoding
\end{itemize}


By recursively partitioning an $\ell \times \ell$ kernel $G_\ell$ into two parts, we construct a binary RMLD tree of depth $\log_2 (\ell) + 1$ (see Fig.~\ref{fig:Combine_RMLD_sections}).  
Each node corresponds to a section $\secc(x\!:\!y)$; non-leaf nodes have two children corresponding to $\secc(x\!:\!z)$ and $\secc(z\!:\!y)$, with $x < z < y$.  
During decoding, the LLR input $L_0^{\ell-1}$ is assigned to the leaf nodes, and each node in $\secc(x\!:\!y)$ produces a soft-output LLR list $T_{xy}$ by combining the outputs of its child nodes, $T_{xz}$ and $T_{zy}$.

Steps 1–3 describe the construction of the RMLD tree, and in Step 4, the input $L_0^{\ell-1}$ is decoded using the RMLD tree.

In Step 1, the kernel $G_\ell$ is transformed into a punctured code $G^p$.  
A punctured code corresponding to $\secc(x\!:\!y)$ is denoted $G^p_{xy}$ and is constructed based on the shortened codes of $\secc(x\!:\!z)$ and $\secc(z\!:\!y)$.  
The code $G^p_{xy}$ is stored in the RMLD node corresponding to $\secc(x\!:\!y)$ and is later used to generate the mapping table in Step 2.

In Step 2, the RMLD node in $\secc(x\!:\!y)$ is connected with its child nodes in $\secc(x\!:\!z)$ and $\secc(z\!:\!y)$.  
To determine the indexing for combining $T_{xz}$ and $T_{zy}$ into $T_{xy}$ during decoding (see \eqref{eq:RMLD_Txy_comb}), a wvab table $\mathbb{T}$ is constructed from $G^p_{xy}$ and the punctured codes of its child nodes, $G^p_{xz}$ and $G^p_{zy}$.

In Step 3, a binary max tree is constructed to perform the maximization in \eqref{eq:RMLD_Txy_comb}.


After Steps 1–3, the RMLD tree has been constructed, with each node containing its corresponding wvab table $\mathbb{T}$ and a max tree to evaluate \eqref{eq:RMLD_Txy_comb}.  
In Step 4, the input LLR $L_0^{\ell-1}$ is assigned to the leaf nodes, and the soft outputs $T$ are propagated from the leaves to the root.  
Each node in $\secc(x\!:\!y)$ generates $T_{xy}$ from its child outputs $T_{xz}$ and $T_{zy}$ using \eqref{eq:RMLD_Txy_comb}.  
Finally, the soft decoding output at phase $i$, $\hat{L}_i$, is obtained from the root node output $T_{0\ell}$ according to \eqref{eq:RMLD_T0l_comb}.

\begin{figure}[ht]
\centering
\centerline{\includegraphics[width=0.6\textwidth]{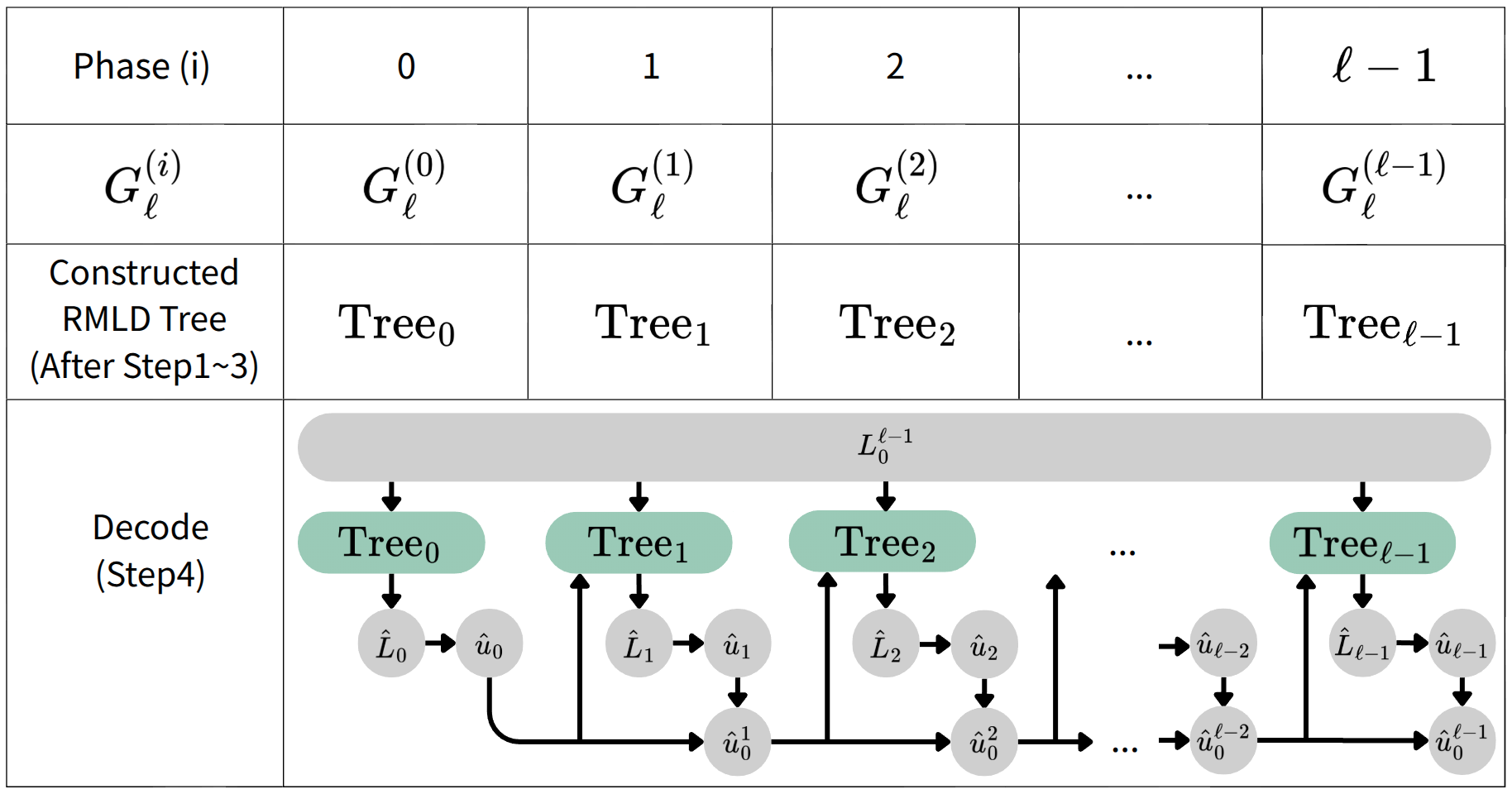}}
\caption{
Overview of the generalized SC decoding for a size-\( \ell \) kernel \( G_\ell \) using the RMLD algorithm. At phase $i$,
 an extended kernel 
 $G_\ell^{(i)}$ is constructed to form an RMLD tree, denoted by $\text{Tree}_i$, following Steps 1-3 in Sec. \ref{Sec:RMLD_steps_}. During decoding, given the received LLRs
 $L^{\ell-1}_0$ and the prior hard decisions $\hat{u}_0^{i-1}$ at phase $i$, a soft LLR output $\hat{L}_i$ is computed using $\text{Tree}_i$ following Step 4. 
  The corresponding hard decision $\hat{u}_i$ is then obtained from $\hat{L}_i$ according to \eqref{eq:hard_decision}. The decoding process proceeds sequentially from phase $0$ to $\ell-1$, ultimately producing the decoded sequence $\hat{u}_0^{\ell-1}$.  }
\label{fig:RMLD_dec_overview}
\end{figure}

In the remainder of this section, we provide examples for each of these steps, although these examples use different kernels to make them easier to understand.



Fig.~\ref{fig:RMLD_dec_overview} illustrates the architecture of the RMLD for decoding a kernel $G_\ell$. At each phase $i$, an RMLD tree (denoted by $\text{Tree}_i$) is constructed from the extended kernel $G^{(i)}_\ell$ according to Steps~1–3. During decoding, $\text{Tree}_i$ takes as input the received LLRs $L_0^{\ell-1}$ together with the prior hard decisions $u_0^{i-1}$, and executes Step~4 to produce the soft output $\hat{L}_i$. Subsequently, a hard decision is made according to \eqref{eq:hard_decision} to obtain $\hat{u}_i$. The estimate $\hat{u}_i$ is then concatenated with $\hat{u}_0^{i-1}$ to form $\hat{u}_0^{i}$, which serves as the prior decision input for the next phase $(i+1)$.

\subsubsection{Punctured Code Construction} 
\label{sec:Punctured Code Construction}
Let us denote by \( G^p_{xy} \) and \( G^s_{xy} \) the generator matrices corresponding to \( p_{xy}(\mathcal{C}) \) and \( s_{xy}(\mathcal{C}) \)-- the punctured and shortened code corresponding to $\mathcal{C}$, respectively.
Suppose the interval \( \secc(x\!:\!y) \) is partitioned into two subintervals, \( \secc(x\!:\!z) \) and \( \secc(z\!:\!y) \), where \( x < z < y \).  In our setting, we define \( z = \left\lfloor \tfrac{x+y}{2} \right\rfloor \).
Then,  \( G^p_{xy} \) can be written as:

\ea{\label{eq:RTPA_matrix}
G^p_{xy} = 
\left[
\begin{array}{*{2}c}
G^{s}_{xz}&0\\
0&G^{s}_{zy}\\
G^{w(0)}_{xy}&G^{w(1)}_{xy}\\
\hline
G^{v(0)}_{xy}&G^{v(1)}_{xy}\\
\end{array}
\right]
=
\left[
\begin{array}{*{1}c}
G^{s}_{xy}\\
\hline
G^{v}_{xy}\\
\end{array}
\right],
}
where \( G^v_{xy} = \left[ G^{v(0)}_{xy}, G^{v(1)}_{xy} \right] \) denotes the generator of the subcode of \( G^p_{xy} \) corresponding to codewords that are not part of the shortened code over \( \secc(x\!:\!y) \). 
Next, let \( G^{w}_{xy} = \left[ G^{w(0)}_{xy}, G^{w(1)}_{xy} \right] \) denote the generator of the subcode of \( G^s_{xy} \) consisting of codewords that do not belong to the shortened codes over either \( \secc(x\!:\!z) \) or \( \secc(z\!:\!y) \). 
%
%
Given a matrix \(G\), we can transform it into \(G^{p}_{xy}\) using the following steps \cite{Tri_RTPA_2021}:  
\begin{itemize}[]
    \item \emph{\underline{Step 1.A}:} To obtain the shortened code \(G^s\), Gaussian elimination is applied to transform \(G\) into a trellis-oriented generator matrix \(\mathbb{G}\) (i.e., a minimum-span generator matrix) \cite{KscTreG95}.   The submatrix \(G^s_{xy}\) consists of the rows of \(\mathbb{G}\) whose entries are zero outside the interval \(\secc(x\!:\!y)\). Using the same procedure, \(G^s_{xz}\) and \(G^s_{zy}\) can also be obtained.
    \item \emph{\underline{Step 1.B}:} The matrix \(G^{v}_{xy}\) is derived from \(\mathbb{G}\) by puncturing outside the interval \(\secc(x\!:\!y)\), yielding \(\mathbb{G}_{xy}\).  
    From \(\mathbb{G}_{xy}\), the rows corresponding to \(G^s_{xy}\) are removed, and the remaining rows are reduced by eliminating linear dependencies.  
    The resulting matrix is denoted as \(G^v_{xy}\).
    \item \emph{\underline{Step 1.C}:} \(G^w_{xy}\) is obtained from \(G^s_{xy}\) by eliminating the rows that also belong to \(G^s_{xz}\) or \(G^s_{zy}\).
\end{itemize}

Next, we provide an example of this procedure.
\begin{example}
Consider $F_4 = F_2^{\otimes 2}$ 
\ea{
F_4 = 
\left[
\begin{array}{cccc}
1&0&0&0\\
1&1&0&0\\
1&0&1&0\\
1&1&1&1\\
\end{array}
\right],
\label{eq:F4}
}
and consider the extended kernel construction in Sec. \ref{sec: Extended Kernel Construction}.
Then, in phase 0, the extended kernel $F_4^{(0)}$ is obtained as  
\ea{
F_4^{(0)} = 
\left[
\begin{array}{ccccc}
1&0&0&0&1\\
1&1&0&0&0\\
1&0&1&0&0\\
1&1&1&1&0\\
\end{array}
\right].
\label{eq:F4(0)}
}

The trellis-oriented generator matrix $\mathbb{G}$ for $F_4^{(0)}$ is 
$$
\mathbb{G} = 
\left[
\begin{array}{ccccc}
1&1&0&0&0\\
0&1&1&0&0\\
0&0&1&1&0\\
0&0&0&1&1\\
\end{array}
\right].
$$
Next we apply the steps 1.A--1.C detailed above.

\smallskip
\noindent
$\bullet$ \emph{\underline{step 1.A}:}
From $\mathbb{G}$
we can obtain \ea{
G^s_{04} = \left[
\begin{array}{ccccc}
1&1&0&0\\
0&1&1&0\\
0&0&1&1\\
\end{array}
\right]}
and
\ea{
G^s_{02}=G^s_{24} =\left[
\begin{array}{ccccc}
1&1\\
\end{array}
\right].
}

\smallskip
\noindent
$\bullet$ \emph{\underline{step 1.B}:} We obtain 
\ea{
G^v_{04} =\left[
\begin{array}{ccccc}
0&0&0&1\\
\end{array},
\right]}
by removing $G^s_{04}$ in $\mathbb{G}$.   

\smallskip
\noindent
$\bullet$ \emph{\underline{step 1.C}:}
Finally, we have that 
\ea{
G^w_{04} = \left[
\begin{array}{ccccc}
0&1&1&0\\
\end{array} 
\right],
}
from $G^s_{04}$.   

\medskip

In our implementation, we modify $G^v$ and $G^w$ by adding selected rows from the original matrix $F^{(0)}_4$, ensuring that all rows in $G^v$ and $G^w$ are drawn from $F^{(0)}_4$. For example, the row $G^v_{04}$ of $F_4^{(0)}$  corresponds to row $\left[
\begin{array}{ccccc}
0&0&0&1&1\\
\end{array} \right],$ which, in the trellis-oriented matrix $\mathbb{G}$ of $F_4^{(0)}$, is given by 
$$
\mathbb{G}[3] = F^{(0)}_4[0]\oplus F^{(0)}_4[1]\oplus F^{(0)}_4[2]\oplus F^{(0)}_4[3].
$$
Similarly, $G^w_{04}$ corresponds to 
$$
\mathbb{G}[1]=F^{(0)}_4[1]\oplus F^{(0)}_4[2].
$$
The shortened codes $G^s_{02}$ and $G^s_{24}$ correspond to 
$$
\mathbb{G}[0]=F^{(0)}_4[1]\space \text{ and } \space \mathbb{G}[2]=F^{(0)}_4[1]+F^{(0)}_4[3],
$$
respectively.  

We eliminate $F_4^{(0)}[1]$ and $F_4^{(0)}[3]$ from the vectors corresponding to $G^v_{04}$ and $G^w_{04}$, since  these vectors form $G^s_{02}$ and $G^s_{24}$. After this elimination, the vector corresponding to $G^v_{04}$ becomes $F^{(0)}_4[0]\oplus F^{(0)}_4[2]$, and the vector corresponding to $G^w_{04}$ becomes $F^{(0)}_4[2]$. Next, we remove $F^{(0)}_4[2]$ from the vector corresponding to $G^v_{04}$, since $F^{(0)}_4[2]$ corresponds to $G^w_{04}$. 

The resulting $G^p_{04}$ for $F^{(0)}_4$ is therefore:
\ea{\label{eq:punc_F4_0}
G^p_{04} = \left[
\begin{array}{c}
G^s_{04}\\
\hline
G^v_{04}\\
\end{array}
\right]=
\left[
\begin{array}{ccccc}
1&1&0&0\\
0&0&1&1\\
1&0&1&0\\
\hline
1&0&0&0\\
\end{array}
\right],
}
where 
$$
G^v_{04}=\left[
\begin{array}{ccccc}
1&0&0&0\\
\end{array} 
\right],\quad  G^w_{04}=\left[
\begin{array}{ccccc}
1&0&1&0\\
\end{array} 
\right],
$$
corresponding to  
$F^{(0)}_4[0]$  and  $F^{(0)}_4[2]$, respectively.
\end{example}
 
\bigskip

The partitioning operation in \eqref{eq:RTPA_matrix} and step 1.A--C  can be applied recursively to \( G^p_{xz} \) and \( G^p_{zy} \).
%
%
%
%
%
More precisely, in phase $i$ of the extended kernel construction in Sec. \ref{sec: Extended Kernel Construction}, an extended kernel $G_\ell^{(i)}$ of size ($\ell-i,\ell+1$) can be punctured to form $G^p_{0\ell}$, as shown in  \eqref{eq:RTPA_matrix}. 
This matrix can then be recursively partitioned into two sections until the section length reduces to one. That is,
\ea{
G^p_{xy}\rightarrow [G^p_{xz},G^p_{zy}], 
\label{eq:recursive split}
}
where $z=\left\lfloor \frac{x+y}{2} \right\rfloor$, and the initial values of $x$ and $y$ are $0$ and $\ell$, respectively.

\subsubsection{Mapping Table Construction}
\label{sec:Mapping Table Construction}

Based on \( G^p_{xy} \), \( G^p_{xz} \), and \( G^p_{zy} \), we construct a binary table—referred to as the \( (w,v,a,b) \) table following the convention in \cite{TriRTPA23} — which captures the mapping relationships among the subcodes \(  G^{w}_{xy}, G^{v}_{xy} \), and \( G^{v}_{xz}, G^{v}_{zy} \). We denote this table by \( \mathbb{T}(w, v, a, b) \).
%
   %

The \( \mathbb{T}(w,v,a,b) \) is constructed as follows, using the expressions in \eqref{eq:RTPA_matrix} by letting \( w \), \( v \), \( s_{xz} \), \( s_{zy} \), \( a \), and \( b \) be binary vectors corresponding to the matrices \( G^w_{xy} \), \( G^v_{xy} \), \( G^s_{xz} \), \( G^s_{zy} \), \( G^v_{xz} \), and \( G^v_{zy} \), respectively.
\eas{
p_{xz}(\mathcal{C})=
\left[
\begin{array}{cc}s'_{xz}, a\end{array}\right]
\left[
\begin{array}{cc}
G^{s}_{xz}\\
\hline G^{v}_{xz}\\
\end{array}\right] = 
\left[
\begin{array}{cc}
s_{xz}, w, v
\end{array}
\right]
\left[
\begin{array}{cc}
G^{s}_{xz}\\ G^{w(0)}_{xy}\\
\hline
G^{v(0)}_{xy}\\
\end{array}
\right]\\
p_{zy}(\mathcal{C}) = 
\left[\begin{array}{cc}s'_{zy}, b\end{array} \right]
\left[\begin{array}{cc} G^{s}_{zy}\\ \hline  G^{v}_{zy}\\ \end{array} \right]
=
\left[\begin{array}{cc}s_{zy}, w, v \end{array} \right]
\left[\begin{array}{cc}
G^{s}_{zy}\\
G^{w(1)}_{xy}\\
\hline
G^{v(1)}_{xy}\\
\end{array} \right].
}{\label{eq:table}}
%
Let \( s'' = s' - s \). By subtracting \( s \) from both sides of the equations in \eqref{eq:table}, the matrix \( G^{s} \) on the right-hand side can be eliminated. 
Subsequently, applying the pseudoinverse of the matrix
 $\left[\begin{array}{cc} G^{s} \\ \hline
G^{v}  \\ \end{array}\right]$
yields the following expressions.

$$
\left[\begin{array}{cc}s''_{xz}, a\end{array}\right]
= 
\left[\begin{array}{cc} w, v \end{array}\right]
\left[\begin{array}{cc} G^{w(0)}_{xy}\\ \hline
G^{v(0)}_{xy}\\ \end{array}\right]
\left[\begin{array}{cc}G^{s}_{xz}\\
\hline G^{v}_{xz}\\
\end{array}\right] ^{-1}
$$
\begin{equation} \label{eq:G_wvab_right}
\left[\begin{array}{cc}s''_{zy}, b\end{array}\right]
=
\left[\begin{array}{cc}w, v \end{array}\right]
\left[\begin{array}{cc} G^{w(1)}_{xy}\\
\hline G^{v(1)}_{xy}\\  
\end{array}\right]\left[\begin{array}{cc} 
G^{s}_{zy}\\ 
\hline  G^{v}_{zy}\\ 
\end{array}\right] ^{-1}.
\end {equation}

To simplify the notation, define
$$
\hat{G}_{xy} = \left[\begin{array}{cc} 
G^{w(0)}_{xy}\\ \hline
G^{v(0)}_{xy}\\ \end{array}\right]
\left[\begin{array}{cc} G^{s}_{xz}\\
\hline G^{v}_{xz}\\
\end{array}\right] ^{-1}
$$
\begin{equation} \label{eq:G_til_def}
\tilde{G}_{xy} =  \left[\begin{array}{cc} 
G^{w(1)}_{xy}\\
\hline G^{v(1)}_{xy}\\  
 \end{array}\right]
 \left[\begin{array}{cc} 
G^{s}_{zy}\\ 
\hline  G^{v}_{zy}\\ 
 \end{array}\right] ^{-1}
\end{equation}
Using \eqref{eq:G_til_def}, $\mathbb{T}(w,v,a,b)$ can then be expressed as
$$
\left[\begin{array}{cc}s''_{xz}, a \end{array}\right]
= \left[\begin{array}{cc} w, v  \end{array}\right] \hat{G}_{xy}
$$
\begin{equation} \label{eq:G_til}
\left[\begin{array}{cc} s''_{zy}, b \end{array}\right]
= \left[\begin{array}{cc}  w, v  \end{array}\right]\tilde{G}_{xy}
\end {equation}

\medskip

\begin{example}\label{ex:wvab_F4}
Consider the $4 \times 4$ Arıkan kernel $F_4$. The extended kernel at phase $i = 1$ is given by
\ea{
F_4^{(1)} = \begin{bmatrix}
1 & 1 & 0 & 0 & 1 \\
1 & 0 & 1 & 0 & 0 \\
1 & 1 & 1 & 1 & 0 \\
\end{bmatrix}.
\label{eq:F4(1)}
}
The corresponding punctured code $G^p_{04}$ for $F_4^{(1)}$, as well as the matrices $\hat{G}$ and $\tilde{G}$ derived from $G^p_{04}$, can be expressed as follows:    

\ea{\label{eq:punc_F4_1}
G^{p}_{04} = 
\left[
\begin{array}{cccc}
  1&0&1&0\\
  1&1&1&1\\
  1&1&0&0
\end{array}
\right]
\leftarrow \! \! \! \lcb \p{w_0\\ w_1\\ v_0} \rnone
, 
G^{p}_{02}= 
\left[
\begin{array}{cc}
  1&0\\
  1&1\\
\end{array}
\right]
\leftarrow \! \! \! \lcb \p{v_0(a_0)\\ v_1(a_1)\\} \rnone
,  
G^{p}_{24}= 
\left[
\begin{array}{cc}
  1&0\\
  1&1\\
\end{array}
\right]
\leftarrow \! \! \! \lcb \p{v_0(b_0)\\ v_1(b_1)\\} \rnone.
}
   
Accordingly, we have
\eas{
\hat{G}=\left[
\begin{array}{cc}
  1&0\\
  1&1\\
  1&1
\end{array}
\right]
\left[
\begin{array}{cc}
  1&0\\
  1&1\\
\end{array}
\right]^{-1} = \left[
\begin{array}{cc}
  1&0\\
  1&1\\
  1&1
\end{array}
\right]
\left[
\begin{array}{cc}
  1&0\\
  1&1\\
\end{array}
\right] =
\left[
\begin{array}{cc}
  1&0\\
  0&1\\
  0&1
\end{array}
\right]
\\
\tilde{G}=
\left[
\begin{array}{cc}
  1&0\\
  1&1\\
  0&0
\end{array}
\right]\left[
\begin{array}{cc}
  1&0\\
  1&1\\
\end{array}
\right]^{-1}
=
\left[
\begin{array}{cc}
  1&0\\
  1&1\\
  0&0
\end{array}
\right]\left[
\begin{array}{cc}
  1&0\\
  1&1\\
\end{array}\right] =\left[\begin{array}{cc}
  1&0\\
  0&1\\
  0&0
\end{array}\right].
}{\label{eq:example 2 G mat}}
The indices $a$ and $b$ (corresponding to the $v$-rows in \( \secc(x\!:\!z) \) and \( \secc(z\!:\!y) \), respectively) can be derived from the indices $w$ and $v$ in the larger section \( \secc(x\!:\!y) \) using the matrices $\hat{G}$ and $\tilde{G}$, as defined in \eqref{eq:G_til}. This yields the transformation table $\mathbb{T}(w,v,a,b)$, as shown in Table~\ref{tab:wvab_F4}.
\end{example}

\begin{table}[htbp]
\centering
\captionsetup{justification=centering}
\caption{The \( \mathbb{T}(w,v,a,b) \) for Example \ref{ex:wvab_F4}.   
The matrix $F_4^{(1)}$ is given in~\eqref{eq:F4(1)}, \\
and $F_2^{(0)}$ is provided in~\eqref{eq:F2_0}.
The entries in the table are obtained from \eqref{eq:G_til}.  
%
}
\centering
\begin{minipage}{0.5\linewidth}
\centering
\begin{tabular}{|c|c|c|c|c|c|c|c|}
\hline
$r$ & $w_0$ & $w_1$ & $v_0$  & $a_0$ & $a_1$ & $b_0$ & $b_1$\\
\hline
0 & 0 & 0 & 0 & 0 & 0 & 0 & 0\\
1 & 0 & 0 & 1 & 0 & 1 & 0 & 0\\
2 & 0 & 1 & 0 & 0 & 1 & 0 & 1\\
3 & 0 & 1 & 1 & 0 & 0 & 0 & 1\\
4 & 1 & 0 & 0 & 1 & 0 & 1 & 0\\
5 & 1 & 0 & 1 & 1 & 1 & 1 & 0\\
6 & 1 & 1 & 0 & 1 & 1 & 1 & 1\\
7 & 1 & 1 & 1 & 1 & 0 & 1 & 1\\
\hline
\end{tabular}
\vspace{2pt}
\caption*{(a) $\mathbb{T}(w,v,a,b)$ for $F_4^{(1)}$ in $\secc(0\!:\!4)$.}
\label{tab:wvab_F4}
\end{minipage}%
\hfill
\begin{minipage}{0.5\linewidth}
\centering
\begin{tabular}{|c|c|c|c|c|}
\hline
$r$ & $w_0$  & $v_0$  & $a_0$ & $b_0$ \\
\hline
0 & 0 & 0 & 0 & 0\\
1 & 0 & 1 & 1 & 0\\
2 & 1 & 0 & 1 & 1\\
3 & 1 & 1 & 0 & 1\\
\hline
\end{tabular}
\vspace{2pt}
\caption*{(b) $\mathbb{T}(w,v,a,b)$ for $F_2^{(0)}$ in $\secc(0\!:\!2)$.}
\end{minipage}
\label{tab:wvab_compare}
\end{table}

\subsubsection{Max Tree Construction}
\label{sec:Max Tree Construction}
%
After obtaining \( \mathbb{T}(w, v, a, b) \), we proceed to construct a set of \emph{max trees}, which are tree structures that enable efficient computation of metrics accumulated along their branches, as will be detailed later.
The total number of max trees  that RMLD constructs is \( 2^{|v|} \). 
The size of each max tree is determined by the number of rows in \( w \) and \( v \), 
denoted as \( |w| \) and \( |v| \), respectively.
Each max tree is a binary tree with depth equal to \( |w| + 1 \).

The node \( \secc(x\!:\!y) \) is associated with a soft-output list -- which we denote as 
\( T_{xy} \). 
The size of \( T_{xy} \) is \( 2^{|v|} \), corresponding to the number of max trees constructed for \( G^{p}_{xy} \).  
%
The node associated with  \( \secc(x\!:\!y) \) takes \( T_{xz} \) and \( T_{zy} \) as inputs and produces \( T_{xy} \) as output. 

Let $\mathbb{T}_r = [w^{(r)}, v^{(r)}, a^{(r)}, b^{(r)}]$ denote the $r$-th row of the table $\mathbb{T}(w,v,a,b)$, where $w^{(r)},v^{(r)},a^{(r)}$ and $b^{(r)}$ are the vectors $w,v,a$ and $b$  in row $\mathbb{T}_r$. 
For a given $v$, define
\ea{
\mathbb{T}^{(v)} = \{\, [a^{(r)}, b^{(r)}] \mid v^{(r)} = v \,\}.
\label{eq:Tv}
}
That is, $\mathbb{T}^{(v)}$ contains all $[a,b]$ pairs from rows $r$ with $v^{(r)} = v$.
Using $\mathbb{T}^{(v)}$ in \eqref{eq:Tv}, the value $T_{xy}[\dec(v)]$ can be computed as
\ea{\label{eq:RMLD_Txy_comb}
T_{xy}[\dec(v)] = \max_{[a,b] \in \mathbb{T}^{(v)}} 
\bigl( T_{xz}[\dec(a)] + T_{zy}[\dec(b)] \bigr).
}
where the function \( \dec(\cdot) \) converts a binary index list into its corresponding decimal index. 
The maximization in \eqref{eq:RMLD_Txy_comb} is implemented using max trees, where each leaf node is assigned a value $N_r$, defined as
\ea{\label{eq:RMLD_Leaf_node_value_Nr}
N_r = T_{xz}[\dec(a^{(r)})] + T_{zy}[\dec(b^{(r)})].
}
%
The leaf node assigned with $N_r$ is determined by the indices $w^{(r)}$ and $v^{(r)}$. 
We first select the max tree corresponding to $v^{(r)}$, then traverse it from the root according to the bits of $w^{(r)}$: 
a 0 directs to the left child, and a 1 to the right. 
The resulting leaf node is assigned the value $N_r$.

\begin{example}    
    For kernel $F_4^{(1)}$ in Example \ref{ex:wvab_F4},
    
\ea{G^{p}_{04} = 
\left[
\begin{array}{cccc}
  1&0&1&0\\
  1&1&1&1\\
  1&1&0&0
\end{array}
\right]
\leftarrow \! \! \! \lcb \p{w_0\\ w_1\\ v_0} \rnone}
    
Since $|w| = 2$ and $|v| = 1$, it follows that there are $2^{|v|} = 2$ max trees, as depicted in Fig.~\ref{fig:MaxTree figure F4_1}, each having a depth of $|w| + 1 = 3$.

In the max tree, the left child node corresponds to index 0, while the right child node corresponds to index 1. For example, in Fig.~\ref{fig:MaxTree figure F4_1}, the first (leftmost) leaf node corresponds to the indices $v_0 = 0$, $w_0 = 0$, and $w_1 = 0$. The third leaf node corresponds to the indices $v_0 = 0$, $w_0 = 1$, and $w_1 = 0$.
The table $\mathbb{T}(w,v,a,b)$ for $F_4^{(1)}$ is provided in Table \ref{tab:wvab_F4}. Each leaf node of the max tree is assigned a value $N_r$ according to \eqref{eq:RMLD_Leaf_node_value_Nr}.   

For example, consider the leaf value $N_5$ in Fig.~\ref{fig:MaxTree figure F4_1}. The 5-th row of $\mathbb{T}(w,v,a,b)$ in Table \ref{tab:wvab_F4} is 
$\mathbb{T}_5 = [1,0,1,1,1,1,0]$. 
Substituting $a^{(5)}=[1,1]$ and $b^{(5)}=[1,0]$ into \eqref{eq:RMLD_Leaf_node_value_Nr} we obtain
\[N_5 = T_{xz}[\dec([1,1])]+T_{zy}[\dec([1,0])]= T_{xz}[3]+T_{zy}[1]\]
$N_5$ is assigned to the position corresponding to \( w^{(5)} = [1,0] \) and \( v^{(5)} = [1] \) in \figref{fig:MaxTree figure}.  
After executing the max trees, an output list \( T_{xy} \) of size \( 2^{|v|} = 2 \) is obtained, where \( T_{xy}[0] \) and \( T_{xy}[1] \) represent the root node values of the left and right trees, respectively.
\end{example}


\subsubsection{RMLD Decoding}
\label{sec:RMLD Decoding}
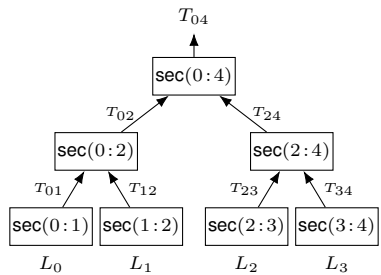
\begin{figure}[htbp]
    \centering

    \begin{subfigure}[b]{0.95\linewidth}
        \centering
        \begin{tikzpicture}[>=Latex, node distance=1.5cm, on grid]
            \tikzstyle{state}=[rectangle, draw, minimum size=5mm, inner sep=1pt]
            \node (T04) at          (   0-2,4.3) {\scriptsize $T_{04}$};
            \node[state] (v0) at   (   0-2,3.5) { \scriptsize $\secc(0\!:\!4)$};
            \node[state] (l0_0) at (  -1.3-2,2.5) {\scriptsize $\secc(0\!:\!2)$};
            \node[state] (l0_1) at (   1.3-2,2.5) {\scriptsize $\secc(2\!:\!4)$};
            \node[state] (l1_0) at (-0.6-1.3-2,1.5) {\scriptsize $\secc(0\!:\!1)$};
            \node[state] (l1_1) at ( 0.6-1.3-2,1.5) {\scriptsize $\secc(1\!:\!2)$};
            \node[state] (l1_2) at (-0.6+1.3-2,1.5) {\scriptsize $\secc(2\!:\!3)$};
            \node[state] (l1_3) at ( 0.6+1.3-2,1.5) {\scriptsize $\secc(3\!:\!4)$};
            \node at               (-0.6-1.3-2,1.0) {\scriptsize $L_0$};
            \node at               ( 0.6-1.3-2,1.0) {\scriptsize $L_1$};
            \node at               (-0.6+1.3-2,1.0) {\scriptsize $L_2$};
            \node at               ( 0.6+1.3-2,1.0) {\scriptsize $L_3$};
            \draw[->] (v0) -- node[] {} (T04);
            \draw[->] (l0_0) -- node[left, font=\tiny] {$T_{02}$} (v0);
            \draw[->] (l0_1) -- node[right, font=\tiny] {$T_{24}$} (v0);
            \draw[->] (l1_0) -- node[left, font=\tiny] {$T_{01}$} (l0_0);
            \draw[->] (l1_1) -- node[right, font=\tiny] {$T_{12}$} (l0_0);
            \draw[->] (l1_2) -- node[left, font=\tiny] {$T_{23}$} (l0_1);
            \draw[->] (l1_3) -- node[right, font=\tiny] {$T_{34}$} (l0_1);
            
        \end{tikzpicture}
    \end{subfigure}
    \captionsetup{justification=centering}
    \caption{Decoding process of the RMLD tree for the size-$4$ kernel $G_4$:\\
      $T_{xy}$ is obtained by combining $T_{xz}$ and $T_{zy}$.\\}
    \label{fig:Combine_RMLD_sections}
\end{figure}

We are finally able to come to the last step of the algorithm and show how the solution of \eqref{eq:llr_sc} is obtained from the quantities and data structures in the steps above.
The RMLD algorithm starts from the root node \( \secc(0\!:\! \ell) \) and  recursively partitions each section \( \secc(x\!:\!y) \) into two subsections: \( \secc(x\!:\!z) \) and \( \secc(z\!:\!y) \). Each section corresponds to an RMLD node, which contains an associated max tree. 
This recursive splitting continues until the section length is equal to one  (i.e. for \(\secc(x\!:\!y), y = x+1\)), at which point a node is made to be a leaf.
During the recursive process, the RMLD nodes collectively form an RMLD tree.
Fig. \ref{fig:Combine_RMLD_sections} illustrates the RMLD tree for the size-4 kernel \( G_4 \). Each RMLD node generates a soft output \( T_{xy} \) from the inputs \( T_{xz} \) and \( T_{zy} \) by performing  max tree updates corresponding to \( \secc(x\!:\!y) \), as defined in \eqref{eq:RMLD_Txy_comb} and \eqref{eq:RMLD_Leaf_node_value_Nr}.

For a leaf node of the RMLD tree corresponding to \( \secc(x\!:\!x+1) \), where the section length is one, the output is set to the LLR values:
\begin{equation}\label{eq:leaf_node_T}
T_{x(x+1)} = [+L_x, -L_x].
\end{equation}
Here, \( L_x \) denotes the \( x \)-th received LLR, as defined in \eqref{eq:llrs}.
When decoding using the RMLD tree, the LLR soft input \( L_0^\ell \) is assigned to the RMLD leaf nodes as described in \eqref{eq:leaf_node_T} and propagated to the RMLD tree root node. Upon reaching the RMLD root node, which corresponds to \( \secc(0\!:\!\ell) \), the soft output \( T_{0\ell} \) is obtained. 
Since the value of \(|v|\) for \( \secc(0\!:\!\ell) \) is always equal to 1, the length of \( T_{0\ell} \) is \( 2^{|v|} = 2 \). 
Subsequently, the following equation is used to compute \( \hat{L}_i \), which is equivalent to the result of \eqref{eq:llr_sc}:
\begin{equation}\label{eq:RMLD_T0l_comb}
  \hat{L}_i = \frac{T_{0\ell}[0] - T_{0\ell}[1]}{2}.
\end{equation}

\begin{figure}[htbp]
        \begin{subfigure}[b]{0.45\linewidth}
        \begin{tikzpicture}[>=Latex, node distance=1.5cm, on grid]

            \node at (0,3.9) {\large rmld\_04};
            \node at               (0,3.4) {\large  $\secc(0\!:\!4)$};
            \node at               (  -2-2,2.5) { $w_0$};
            \node at               (-2.5-2,1.5) { $w_1$};
            \node at               (   0-2,3.5) {$T_{04}[0]$};
            \node at               (  -1.5-2,3) { $v_0$};
            \node[state] (v0) at   (   0-2,3) {0};
            \node[state] (l0_0) at (  -1-2,2.5) {0};
            \node[state] (l0_1) at (   1-2,2.5) {1};
            \node[state] (l1_0) at (-1.6-2,1.5) {0};
            \node[state] (l1_1) at (-0.4-2,1.5) {1};
            \node[state] (l1_2) at ( 0.4-2,1.5) {0};
            \node[state] (l1_3) at ( 1.6-2,1.5) {1};
            \node at               (-1.6-2,1.0) {\scriptsize $N_0$};
            \node at               (-0.4-2,1.0) {\scriptsize $N_2$};
            \node at               ( 0.4-2,1.0) {\scriptsize $N_4$};
            \node at               ( 1.6-2,1.0) {\scriptsize $N_6$};
            \node at               (   0+2,3.5) {$T_{04}[1]$};
            \node[state] (v1) at   (   0+2,3) {1};
            \node[state] (l0_2) at (  -1+2,2.5) {0};
            \node[state] (l0_3) at (   1+2,2.5) {1};
            \node[state] (l1_4) at (-1.6+2,1.5) {0};
            \node[state] (l1_5) at (-0.4+2,1.5) {1};
            \node[state] (l1_6) at ( 0.4+2,1.5) {0};
            \node[state] (l1_7) at ( 1.6+2,1.5) {1};
            \node at               (-1.6+2,1.0) {\scriptsize $N_1$};
            \node at               (-0.4+2,1.0) {\scriptsize $N_3$};
            \node at               ( 0.4+2,1.0) {\scriptsize $N_5$};
            \node at               ( 1.6+2,1.0) {\scriptsize $N_7$};
            
            \draw[->] (l0_0) -- node[above left]  {} (v0);
            \draw[->] (l0_1) -- node[above right] {} (v0);
            \draw[->] (l1_0) -- node[above] {} (l0_0);
            \draw[->] (l1_1) -- node[above] {} (l0_0);
            \draw[->] (l1_2) -- node[above] {} (l0_1);
            \draw[->] (l1_3) -- node[above] {} (l0_1);
            \draw[->] (l0_2) -- node[above left] {} (v1);
            \draw[->] (l0_3) -- node[above right] {} (v1);
            \draw[->] (l1_4) -- node[above] {} (l0_2);
            \draw[->] (l1_5) -- node[above] {} (l0_2);
            \draw[->] (l1_6) -- node[above] {} (l0_3);
            \draw[->] (l1_7) -- node[above] {} (l0_3);
        \coordinate (rect-south-west) at (-5,0.5);  
        \coordinate (rect-north-east) at (5,4.2);      
        \node (recttop)               at (0,4.1)  {};
        \draw[thick, black, rounded corners=0pt] 
            (rect-south-west) rectangle (rect-north-east);
            \node (T04point)    at       (0,4.8) {\large $T_{04}$};
            
            \node (T02point)    at    (-2-2,-0.5)  {\large rmld\_02};
            \node (rectbotleft) at    (-2-1,0.62)  {};
            \node (T24point)    at    ( 2+2,-0.5)  {\large rmld\_24};
            \node (rectbotright)at    ( 2+1,0.62)  {};
            \draw[->] (T02point) -- node[left,font=\large] {$T_{02}$} (rectbotleft);
            \draw[->] (T24point) -- node[right,font=\large] {$T_{24}$} (rectbotright);
            \draw[->] (recttop) -- node[left] {} (T04point);
        \end{tikzpicture}
        \caption{Max tree over \( \secc(0\!:\!4) \) for \( F_4^{(1)} \): The outputs \( T_{02} \) and \( T_{24} \) are combined to produce \( T_{04} \). The leaf values \( N_r \) are computed from \( T_{02} \) and \( T_{24} \) based on \eqref{eq:RMLD_Leaf_node_value_Nr} and $\mathbb{T}_r$ in Table~\ref{tab:wvab_F4}\,(a).}
    \label{fig:MaxTree figure F4_1}
    \end{subfigure}
    \hspace{0.14\textwidth}
    \begin{subfigure}[b]{0.45\linewidth}
        \begin{tikzpicture}[>=Latex, node distance=1.5cm, on grid]
            \node at               (    -2,3.9) {\large rmld\_02};
            \node at               (    -2,3.4) {\large  $\secc(0\!:\!2)$};
            \node at               (  -2-2,2.5) { $v_0$};
            \node at               (-2.5-2,1.5) { $w_0$};
            \node at               (   0-3,3) {$T_{02}[0]$};
            \node at               (   1-2,3) {$T_{02}[1]$};
            \node[state] (l0_0) at (  -1-2,2.5) {0};
            \node[state] (l0_1) at (   1-2,2.5) {1};
            \node[state] (l1_0) at (-1.6-2,1.5) {0};
            \node[state] (l1_1) at (-0.4-2,1.5) {1};
            \node[state] (l1_2) at ( 0.4-2,1.5) {0};
            \node[state] (l1_3) at ( 1.6-2,1.5) {1};
            \node at               (-1.6-2,1.0) {\scriptsize $N_0$};
            \node at               (-0.4-2,1.0) {\scriptsize $N_2$};
            \node at               ( 0.4-2,1.0) {\scriptsize $N_1$};
            \node at               ( 1.6-2,1.0) {\scriptsize $N_3$};
            \draw[->] (l1_0) -- node[above] {} (l0_0);
            \draw[->] (l1_1) -- node[above] {} (l0_0);
            \draw[->] (l1_2) -- node[above] {} (l0_1);
            \draw[->] (l1_3) -- node[above] {} (l0_1);
        \coordinate (rect-south-west) at (-5,0.5);  
        \coordinate (rect-north-east) at (1,4.2);      
        \node (recttop)               at (-2,4.1)  {};
        \draw[thick, black, rounded corners=0pt] 
            (rect-south-west) rectangle (rect-north-east);
            \node (T04point)    at       (-2,4.8) {\large $T_{02}$};
            \node (T02point)    at    (-2-2,-0.5)  {\large rmld\_01};
            \node (rectbotleft) at    (-1-2,0.62)  {};
            \node (T24point)    at    ( 2-2,-0.5)  {\large rmld\_12};
            \node (rectbotright)at    ( 1-2,0.62)  {};
            \draw[->] (T02point) -- node[left,font=\large] {$T_{01}$} (rectbotleft);
            \draw[->] (T24point) -- node[right,font=\large] {$T_{12}$} (rectbotright);
            \draw[->] (recttop) -- node[left] {} (T04point);
        \end{tikzpicture}
        \caption{Max tree over \(\secc\)(0:2)  for $F_2^{(0)}$: The leaf values $N_r$ are computed from $T_{01}$ and $T_{12}$ based on \eqref{eq:RMLD_Leaf_node_value_Nr} and $\mathbb{T}_r$ in Table~\ref{tab:wvab_F4}\,(b).}\hspace{0.3cm}
        \label{fig:MaxTree figure F2}
    \end{subfigure}
     \captionsetup{justification=centering}
    \caption{ RMLD max tree examples.}
    \label{fig:MaxTree figure}
\end{figure}
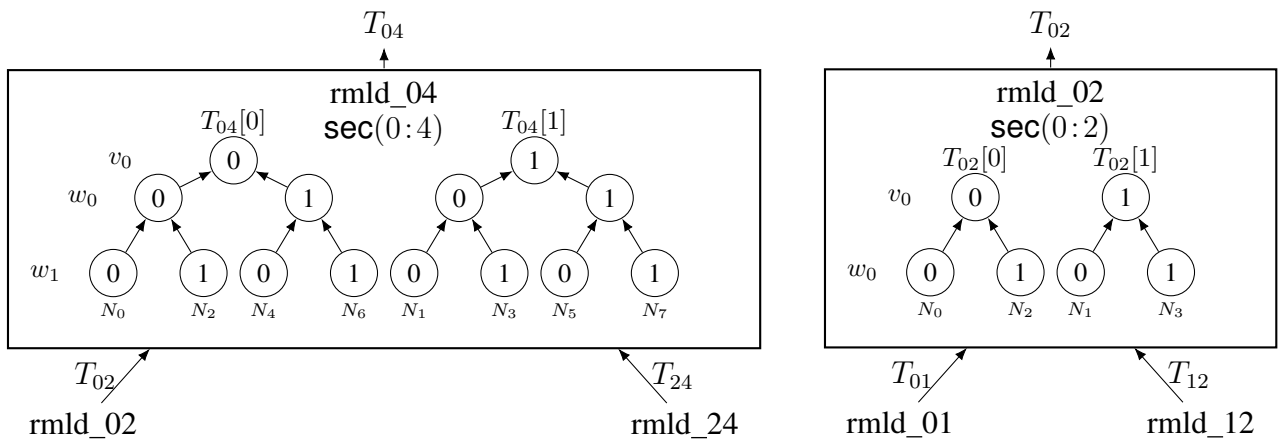

\begin{example}
    RMLD decoding for Arıkan's $F_2$ kernel: In phase 0, the extended kernel is given by $F_2^{(0)}=\left[
\begin{array}{ccc}
  1&0&1\\
  1&1&0\\
\end{array}
\right]$.   

The corresponding punctured codes are $G^p_{02} = \left[
\begin{array}{ccc}
  1&1\\
  1&0\\
\end{array}
\right]\begin{array}{ccc}
  w_0\\
  v_0\\
\end{array}, G^p_{01}=\left[
\begin{array}{ccc}
  1
\end{array}
\right]
\begin{array}{ccc}
  v_0(a_0)
\end{array}, G^p_{12}=\left[
\begin{array}{ccc}
  1
\end{array}
\right]
\begin{array}{ccc}
  v_0(b_0)
\end{array}$. 
The \( \mathbb{T}(w, v, a, b) \) corresponding to \( F_2^{(0)} \) is provided in Table~\ref{tab:wvab_F4}, and the associated max tree structure is illustrated in Figure~\ref{fig:MaxTree figure F2}.

From \eqref{eq:RMLD_Txy_comb}, \eqref{eq:RMLD_Leaf_node_value_Nr} and \eqref{eq:leaf_node_T}, we obtain:
\eas{
N_0 & = T_{01}[0]+T_{12}[0]= +L_0+L_1,\\
N_1 & = T_{01}[1]+T_{12}[0]= -L_0+L_1,\\
N_2 & = T_{01}[1]+T_{12}[1]= -L_0-L_1,\\
N_3 & = T_{01}[0]+T_{12}[1]= +L_0-L_1.
}{\label{eq: Ns}}
The values at the root of the max tree are then given by:
\eas{
T_{02}[0] & =\max(N_0,N_2) =\max(L_0+L_1,-L_0-L_1) = |L_0+L_1|, \\
T_{02}[1] & =\max(N_1,N_3) =\max(-L_0+L_1,L_0-L_1) = |L_0-L_1|.
}{\label{eq:T_02}}
The final soft output is computed as:
\ea{
\hat{L}_0 = \f {+T_{02}[0]-T_{02}[1]} 2 =  \f {|L_0+L_1|-|L_0-L_1|} 2,
}
which corresponds to the well-known \( F \)-function approximation in \eqref{f_func_approx}.
\end{example}

\bigskip
\noindent
Before proceeding further, let us provide a summary of the notation introduced so far in Table \ref{tab:sc_notation}.

\begin{table}[htbp]
\centering
\caption{Summary of notation introduced is Sec. \ref{sec:Recursive Maximum Likelihood Decoding}. }
\label{tab:sc_notation}

\begin{tabular}{|p{2cm}|p{4cm}||p{2cm}|p{4cm}|}
\hline
\textbf{Symbol} & \textbf{Description} & \textbf{Symbol} &  \textbf{Description}\\
\hline
$\ell$ & Kernel size & $G_\ell$ & Kernel with size $\ell \times \ell$ \\
\hline
$F_2$ & Arıkan kernel & $l_i$ & $i$-th bit likelihood ratio:
$l_i\!=\!\exp(L_i)$ \\
\hline
$L_i$ & \( i \)-th input LLR corresponding to the received symbols  & $\hat{L}_i$ & \( i \)-th soft-decision output LLR\\
\hline
$u_i$ & \( i \)-th source bit 
& $\hat{u}_i$ & Decoded estimate of \( u_i \) (hard-decision bit) \\
\hline
$G_\ell^{(i)}$ & Extended generator matrix for decoding phase \( i \) & $\mathbb{G}$ & Trellis-oriented generator
matrix \\
\hline
$\bar{c}^{(i)}$ & Codeword vector in the extended codebook defined by $G_\ell^{(i)}$ & 
$F(L_j\!\mid\!\bar{c}^{(i)}_j\!=\! 0)$ & Selective LLR contribution: $L_j$ if \( \bar{c}^{(i)}_j = 0 \), zero otherwise \\
 \hline 
 $\secc(x\!:\!y)$ & Section in position $[x,y)$ &
$\mathcal{C}_{xy}$ & Subcode of $\mathcal{C}$, such that
all its codewords have non-zero symbols only in $\secc(x\!:\!y)$\\
\hline
$p_{xy}(\mathcal{C})$ &  Linear code obtained by puncturing all symbols outside $\secc(x\!:\!y)$ from codewords of $\mathcal{C}$ & 
$s_{xy}(\mathcal{C})$ & Shortened code of \( \mathcal{C} \), defined as \( p_{xy}(\mathcal{C}_{xy}) \) \\
\hline
$G^{p}_{xy}$ & Punctured code in $\secc(x\!:\!y)$, generator matrix for $p_{xy}(\mathcal{C})$ &
$G^{s}_{xy}$ & Shortened code in $\secc(x\!:\!y)$, generator matrix for $s_{xy}(\mathcal{C})$\\
 \hline 
$w,v$ & Indices corresponding to rows $G^{w}_{xy}$ and $G^{v}_{xy}$ in \eqref{eq:RTPA_matrix} &
$a,b$ & In $\secc(x\!:\!y)$, $a,b$ correspond to $v$ index in $\secc(x\!:\!z)$ and $\secc(z\!:\!y)$, respectively \\
\hline
$\mathbb{T}(w,v,a,b)$ & $(w,v,a,b)$ table that satisfies \eqref{eq:G_til}&
$T_{xy}$ & Output list for a RMLD node in  $\secc(x\!:\!y)$ \\
\hline
\end{tabular}
\end{table}


\section{Problem Formulation}
\label{sec:Problem Formulation}
The original motivation behind the design of the RMLD algorithm lies in reducing computational complexity.
We now proceed to derive the computational complexity of the RMLD algorithm as introduced in Sec. \ref{sec:Recursive Maximum Likelihood Decoding}. 

As described in Sec.~\ref{sec:Recursive Maximum Likelihood Decoding}, the RMLD algorithm adopts a divide-and-conquer strategy: to decode \( \textsf{sec}(x\!:\!y) \), it recursively decodes the sub-sections \( \textsf{sec}(x\!:\!z) \) and \( \textsf{sec}(z\!:\!y) \), and subsequently combines the results.
The complexity of this combination step depends on the number of rows in the matrices \( G^w_{xy} \) and \( G^v_{xy} \) discussed in Sec. \ref{sec:Punctured Code Construction}.  
Let us denote the complexity of this step as \( \Csf_{\text{comb}}(G^p_{xy}) \), which is given by: 
\begin{equation}
\Csf_{\text{comb}}(G^p_{xy}) = 2^{|w|+|v|} + \sum_{k=1}^{|w|} 2^k,
\label{eq:C comb}
\end{equation}
where \(|w|\) and \(|v|\) denote the number of rows in \( G^w_{xy} \) and \( G^v_{xy} \), respectively. 
The complexity is evaluated in terms of the total number of summations and comparisons, as in \cite{TriRTPA23}.

The term $2^{|w|+|v|}$ in \eqref{eq:C comb} accounts for the number of summations required to combine \( \textsf{sec}(x\!:\!z) \) and \( \textsf{sec}(z\!:\!y) \), which corresponds to the number of rows in $\mathbb{T}(w,v,a,b)$ -- see  Table \ref{tab:wvab_F4}.
The term $\sum_{k=1}^{|w|} 2^k $ in \eqref{eq:C comb}  represents the number of comparisons required to execute the max tree -- see 
Fig. \ref{fig:MaxTree figure}.

The total complexity for decoding \( \textsf{sec}(x\!:\!y) \) is thus the sum of the complexities of the two subsections and the combination step:
\begin{equation}
\Csf(G^p_{xy}) 
= \Csf(G^p_{xz}) + \Csf(G^p_{zy}) + \Csf_{\text{comb}}(G^p_{xy}).
\label{eq:comp 2}
\end{equation}
%
%
For a kernel $G_\ell$, there are $\ell$ extended kernels $G^{(i)}_\ell$, where $i\in[\ell]$.  
Define the punctured code corresponding to $G^{(i)}_\ell$ as $G^{p(i)}_{0\ell}$; then the total complexity of computing the soft decoding output  $\hat{L}_i^{\ell-1}$ is given by:

\begin{equation}
\Csf(G_\ell) =  \sum_{i=0}^{\ell-1}( \Csf(G^{p(i)}_{0\ell})+1).
\label{eq:rmld_tot_comp}
\end{equation}
Note that the \( +1 \) term in \eqref{eq:rmld_tot_comp} reflects the summation needed to compute~\eqref{eq:RMLD_T0l_comb} at each decoding phase.

\begin{example}
\label{ex:complexity}
For $F_4 = F_2^{\otimes 2}$,  the extended kernel in phase-0 is given by  $F_4^{(0)}$ in \eqref{eq:F4(0)}.
%
The punctured codes  $G^p_{04}$, $G^p_{02}$, and $G^p_{24}$  for $F_4^{(0)}$ are given as follows:
$$
G^{p}_{04} = \left[
\begin{array}{ccccc}
1&1&0&0\\
0&0&1&1\\
1&0&1&0\\
\hline
1&0&0&0\\
\end{array}
\right]
\leftarrow \! \! \! \lcb \p{s_{02}\\s_{24}\\w_0\\v_0\\} \rnone
, G^{p}_{02} = \left[\begin{array}{ccccc}
1&1\\
\hline
1&0\\
\end{array}\right] 
\leftarrow \! \! \! \lcb \p{w_0\\v_0\\} \rnone
,G^{p}_{24} = \left[\begin{array}{ccccc}
1&1\\
\hline
1&0\\
\end{array}\right] 
\leftarrow \! \! \! \lcb \p{w_0\\v_0\\} \rnone
$$  
All \( \textsf{sec}(0\!:\!4) \), \( \textsf{sec}(0\!:\!2) \) and \( \textsf{sec}(2\!:\!4) \) have $|w|=1$ and $|v|=1$;  thus, the complexity in phase-1 can be calculated using \eqref{eq:comp 2}:
\begin{align*}
 \mathsf{C}(G^p_{02}) & = \mathsf{C}(G^p_{24})   = 2^{2} + \sum_{k=1}^{1} 2^k = 6,\\
  \mathsf{C}(G^p_{04}) & = \mathsf{C}(G^p_{02}) + \mathsf{C}(G^p_{24}) +  2^{2} + \sum_{k=1}^{1} 2^k = 18. 
\end{align*}
\end{example}

In \tabref{tab:compR4F4}, we compare the complexity of Arıkan's kernel $F_4$ in \eqref{eq:F4} with that of the sorted Arıkan's kernel $S_4$ -- see Table \ref{tab:sorted} in Appendix \ref{app:Tabulated Large Polarization Kernels}.
%
%
It can be observed that the total complexity of $F_4$ is 
$\sum_{i=0}^{\ell-1}(\Csf(F^{p(i)}_{0\ell}) + 1) = 64$, whereas the total complexity of $S_4$ is $52$, indicating a reduction in complexity when using the sorted kernel. 

\begin{table}[htbp]
\captionsetup{justification=centering}
\caption{RMLD Complexity for Arıkan's kernel $G_4$ and sorted Arıkan's Kernel $S_4$,\\
$|w_{xy}|$ and $|v_{xy}|$ denotes $|w|$ and $|v|$ for \( \textsf{sec}(x\!:\!y) \), respectively.}
\begin{center}
\begin{tabular}{|c|c|c|c|c|c|}
\hline
  Kernel & $\Csf(G^p_{04})$ & $|w_{04}|,|v_{04}|$  &
   $|w_{02}|,|v_{02}|$ & $|w_{24}|,|v_{24}|$\\
\hline
  $F_4^{(0)}$ & 18 & 1,1 & 1,1 & 1,1\\
  $F_4^{(1)}$ & 22 & 2,1 & 0,2 & 0,2\\ 
  $F_4^{(2)}$ & 14 & 1,1 & 0,2 & 0,2\\ 
  $F_4^{(3)}$ &  6 & 0,1 & 0,1 & 0,1\\ 
\hline
  $S_4^{(0)}$ & 18 & 1,1 & 1,1 & 1,1\\
  $S_4^{(1)}$ & 14 & 0,1 & 1,1 & 1,1\\ 
  $S_4^{(2)}$ & 10 & 1,1 & 0,1 & 0,1\\ 
  $S_4^{(3)}$ &  6 & 0,1 & 0,1 & 0,1\\ 
\hline
\end{tabular}
\label{tab:compR4F4}
\end{center}
\end{table}

\subsection{Trellis reuse} \label{sec:trellis_reuse} 
The decoding complexity can be reduced further by reusing max-trees across decoding phases. Given two decoding phases \( i < j \),  consider  \( \textsf{sec}(x\!:\!y) \) that appears in the max-trees of both phases. If:
\begin{enumerate}
    \item the shortened codes \( G^s_{xz} \) and \( G^s_{zy} \) are identical in both phases \( i \) and \( j \), and
    \item the matrices \( G^w_{xy} \) and \( G^v_{xy} \) in phase \( j \) are subcodes of those in phase \( i \),
\end{enumerate}
then the max-trees constructed for \( \textsf{sec}(x\!:\!y) \)  in phase \( i \) can be reused in phase \( j \), thereby reducing the computational cost of phase \( j \) to one.

By reusing the max-trees across multiple phases, the need to execute the max trees at each phase is eliminated, leading to significant complexity reduction.
Further reductions in complexity are possible by using the special-case optimization techniques proposed in~\cite{TriRTPA23}. However, for simplicity, such techniques are not applied in this work.
Additionally, we only reuse the max-trees corresponding to the largest section \( \textsf{sec}(0\!:\!\ell) \)  of the kernel $G_\ell$. Nonetheless, the proposed search algorithm is general and can accommodate various complexity optimization techniques.

\begin{example}
For the kernel $F_4$ in \eqref{eq:F4}, the punctured code $G^{p(i)}_{04}$  in each decoding phase is given as follows:
\begin{equation*}
G^{p(0)}_{04} = 
\left[
\begin{array}{c c c c}
1&1&0&0\\
0&0&1&1\\
1&0&1&0\\
1&0&0&0\\
\end{array}
\right]
\leftarrow \! \! \! \lcb \p{s_{02}\\ s_{24}\\ w_0\\ v_0\\} \rnone
\end{equation*}

$$
G^{p(1)}_{04} = 
\left[
\begin{array}{c c c c}
1&0&1&0\\
1&1&1&1\\
1&1&0&0\\
\end{array}
\right]
\leftarrow \! \! \! \lcb \p{w_0\\w_1\\v_0\\} \rnone
$$
$$
G^{p(2)}_{04} = 
\left[
\begin{array}{c c c c}
1&1&1&1\\
1&0&1&0\\
\end{array}
\right]
\leftarrow \! \! \! \lcb \p{w_0\\v_0\\} \rnone
$$
$$
G^{p(3)}_{04} = 
\left[
\begin{array}{c c c c}
1&1&1&1\\
\end{array}
\right]
\leftarrow \! \! \! \lcb \p{v_0\\} \rnone
$$
Note that $G^{p(0)}_{04}$ and $G^{p(1)}_{04}$ are the same matrices as those obtained in~\eqref{eq:punc_F4_0} and~\eqref{eq:punc_F4_1}, respectively.
In phase 1, reuse from phase 0 is not possible because the shortened codes $G^s_{02}$ and $G^s_{24}$  used in phase 0 are not present in phase 1.
Both phase 2 and phase 3 can reuse the max-trees in \( \textsf{sec}(0\!:\!4) \) from phase 1, as they do not involve any shortened code $G^s_{02}$ and $G^s_{24}$, and the $w$ and $v$ rows in phase 2 and 3 are subcodes of the $w$ and $v$ rows in phase 1. By reusing the max-trees in \( \textsf{sec}(0\!:\!4) \), the decoding complexity can be reduced from 64 to 44.
  
Similarly, for the sorted Arıkan kernel \( S_4 \), reuse is possible from phase 0 to phase 1 and from phase 2 to phase 3. In this case, reusing the max-trees in \( \textsf{sec}(0\!:\!4) \) reduces the decoding complexity from 52 to 32.
\end{example}

\subsection{Low-complexity RMLD Kernel Search}

Having specified the design criteria in terms of (i) the PDP, as discussed in \secref{sec:Error exponent for large kernel}, and (ii) the decoding complexity under RMLD, as detailed in \secref{sec:Recursive Maximum Likelihood Decoding}, we are now ready to define our design objective formally.

Our goal is to design a kernel \( K \) that exhibits low complexity under RMLD decoding while also approaching the PDP upper bound described in \secref{sec:PDP}.
Formally, the optimization problem is:
\ea{
K^* & = \argmin_{K \in \mathbb{F}_2^{\ell \times \ell}: \: \Dv(K) =  \widetilde{\Dv}(\ell)} \Csf (K),
\label{eq:K minimize C}
}
that is we wish to determine the kernel meeting the desired PDP while attaining the minimum decoding complexity. 
%
%
We refer to the optimization problem in \eqref{eq:K minimize C} as the \emph{Low-complexity RMLD kernel search problem}.
In the next section, we propose a solution to \eqref{eq:K minimize C} based on a reinforcement learning approach -- \algo. This choice is motivated by the fact that, for sufficiently large \( \ell \), handcrafted kernel designs—such as those in \cite{TriTre19,tri64}—are no longer effective due to the exponential growth of the search space.

\section{Reinforcement Learning (RL) Refresher}
\label{sec:RL_refresh}
In this section, we briefly introduce some of the RL agents that will be  relevant in the discussion of the proposed solution -- \algo -- in Sec. \ref{sec:Proposed Approach}.

\subsection{Markov decision process (MDP)}
   
In a Markov Decision Process (MDP)\cite{bellman1957dynamic}\cite{DynamicProgramming2007}, we consider tuples of states ($s$), actions ($a$), and rewards ($r$). At a given state $s$, selecting an action $a$ leads to a transition to a next state $s'$ with probability 
$\Pr(s'|s,a)$, where $\Pr(s'|s,a)$ is referred to as the transition probability. Upon transitioning from $s$ to $s'$ under action $a$, the agent receives a reward denoted by $r(s, a, s')$.  

Formally, an MDP is defined by the tuple $(\mathcal{S}, \mathcal{A}, P, R, \gamma)$, where $\mathcal{S}$ is the state space, $\mathcal{A}$ is the action space, $P$ specifies the transition dynamics, $R$ defines the reward function, and $\gamma \in [0,1)$ is the discount factor that balances immediate and future rewards.  

The objective in an MDP is to find a \textit{policy} $\boldsymbol{\pi}(a|s)$, which specifies the probability of selecting action $a$ in state $s$, that maximizes the expected cumulative discounted reward:
\[
\max_{\boldsymbol{\pi}} \; \mathbb{E}_{\boldsymbol{\pi}} \left[ \sum_{t=0}^{\infty} \gamma^t r(s_t, a_t, s_{t+1}) \right].
\]

\subsection{Value Iteration Algorithm} 
Given a MDP, the value iteration algorithm estimates the future return using the Bellman equation 
%
\begin{equation}\label{eq:bellman}
V'(s) = \max_{a \in A(s)} \sum_{s' \in S} \Pr(s'|s,a) \big[r(s,a,s') + \gamma V(s')\big]
\end{equation}
%
Since in our kernel searching environment the state transition is always deterministic, we can rewrite \eqref{eq:bellman} as  
$$
V'(s) = \max_{a\in A(s)} [r(s,a,s')+\gamma V(s')].
$$

However, the value iteration algorithm needs to store every state value $V(s)$ in the state table, which makes it impractical when the state number is large.   
In our kernel search environment, the number of states is equal to the number of all possible combinations of the kernel that satisfies a given PDP.  

\subsection{Random Agent}
\label{sec:Random Agent}

A Random Agent is a simple baseline in reinforcement learning that selects actions uniformly at random from the set of available actions $A(s)$ at each state $s$. Formally, the policy of a random agent is

\[
\boldsymbol{\pi}_\text{rand}(a \mid s) = \frac{1}{|A(s)|}, \quad \forall \ a \in A(s).
\]

The random agent serves as a baseline to evaluate the performance of more sophisticated reinforcement learning algorithms. 

\begin{figure}[h!]
\centering
\centerline{\includegraphics[width=0.6\textwidth]{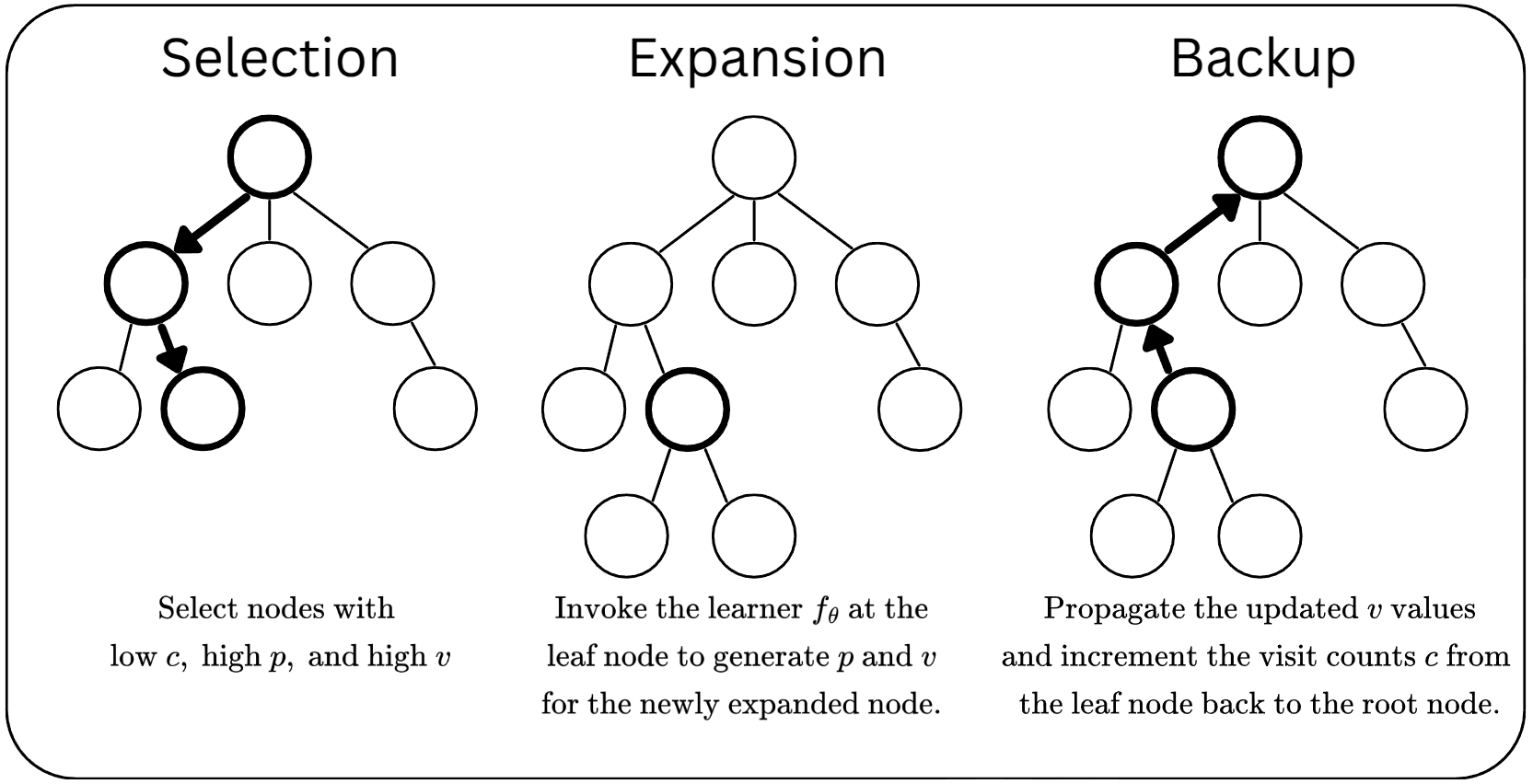}}
\caption{ 
Example of the simulation process in MCTS. At each simulation, MCTS proceeds through the Selection, Expansion, and Backup phases. During the Selection phase, nodes are chosen based on lower visit counts $c$, higher action probabilities $p$, and higher value estimates $v$. In the Expansion phase, the network $f_\theta$ generates the action probabilities $p$ and the value estimate $v$ at the leaf node. These values are then propagated back to the root during the Backup phase, progressively refining the policy $\boldsymbol{\pi}$.
}
\label{fig:MCTS_overview}
\end{figure}

\subsection{Alpha Zero}
\label{sec:Alpha Zero} 

AlphaZero \cite{alphazero} is a self-play RL agent that has mastered various board games without relying on human expertise. 
 The AlphaZero framework consists of two main phases: a self-play phase and a neural network training phase. During the self-play phase, the agent selects actions using the Monte Carlo Tree Search (MCTS) algorithm. The data collected from self-play is then used to train the neural network.
 To accelerate training, we adopt the Gumbel AlphaZero algorithm \cite{gumble_az}, which reduces the search budget required during the self-play phase.

%
Given a RL environment at time step $t$ that takes a discrete action \(a_t\) and state $s_t$ as input and returns a state \(s_{t+1}\) and a reward \(r_t\), 
AlphaZero~\cite{alphazero} employs a neural network 
\((\mathbf{p}_t, v_t) = f_\theta(s_t)\) with parameters \(\theta\) to produce both a policy vector \(\mathbf{p}_t\) and a scalar value \(v_t\). 
Each element of $\mathbf{p}_t$ represents the probability of selecting action $a$ in state $s_t$ as predicted by the neural network. 
Let \(z\) denote the game outcome, which may represent, for example, a win/loss indicator in a board game or the cumulative reward over an episode. 
If \(r_k \in \mathbb{R}\) is the immediate reward obtained at time step \(k\), then  
$z_t = \sum_{k=t}^{T-1} r_k$, where \(T\) is the number of steps in the episode.
The network’s value estimate satisfies $v_t \approx \mathbb{E}[z_t|s_t]$.

AlphaZero uses MCTS to simulate future states \(s\) and values \(v\). 
The MCTS algorithm consists of three main phases: Selection, Expansion, and Backup.
In the search tree, each node represents a state ($s$). Each node stores a visit count $c$, an action probability $p$ and a value estimate $v$. 
\begin{itemize}[]
    \item \emph{Selection:} The selection process starts from the root node and proceeds down to a leaf node. At each step, the selection policy balances exploitation and exploration by prioritizing actions associated with lower visit counts $c$, higher action probabilities $p$, and higher value estimates $v$. 
    \item \emph{Expansion:} After reaching a leaf node, the neural network \(f_\theta\) is evaluated to produce \((\mathbf{p}, v)\). The leaf node is then expanded by adding new child nodes corresponding to the possible next states resulting from each action. 
    \item \emph{Backup:} In the backup phase, the value $v$ obtained from $f_\theta$ is propagated back from the leaf node to the root node. The statistics of all nodes along the traversal path are updated accordingly, typically by incrementing visit counts $c$ and averaging value estimates $v$.
\end{itemize}

The Selection, Expansion, and Backup phases are repeated multiple times until the allocated search budget is exhausted.
After the search process, a policy vector \(\boldsymbol{\pi}\) is derived from the visit counts $c$ at the root node of the search tree. 
An action \(a\) is then sampled from \(\boldsymbol{\pi}\) and executed by the AlphaZero agent.  
Fig. \ref{fig:MCTS_overview} illustrates an example of the simulation process in MCTS.

During training, the network parameters \(\theta\) are optimized using a loss function that combines:  
\begin{enumerate}
    \item a mean squared error (MSE) term between the network's value output \(v\) and the empirical return \(z\),  
    \item a cross-entropy loss between the network’s policy output \(\mathbf{p}\) and the MCTS-derived policy \(\boldsymbol{\pi}\), and  
    \item an \(L_2\) regularization term on \(\theta\).  
\end{enumerate}

The loss function is given by  
\begin{equation}\label{eqn:MCTS_loss}
l = \left( z - v \right)^2 - \boldsymbol{\pi}^\mathsf{T} \log \mathbf{p} + \lambda\lVert \theta \rVert^2 
\end{equation}
where \(\lambda\) is the regularization coefficient.

In~\cite{gumble_az}, it was shown that Gumbel AlphaZero (GAZ) can surpass the original AlphaZero in scenarios with a limited number of search simulations. The core idea behind GAZ is to sample actions without replacement and to apply the sequential halving algorithm at the root node, thereby enhancing search efficiency. In our work, we adopt GAZ based on the implementation provided in~\cite{cpp_minizero}.

\vspace{1cm}

\section{Proposed Approach: \algo}
\label{sec:Proposed Approach}

The proposed approach to the design of low-complexity RMLD kernels—termed \algo—leverages the AlphaZero RL framework \cite{alphazero} to efficiently search for polarization kernels that meet a target PDP while minimizing decoding complexity. 
At a high level, \algo \space operates by iteratively constructing a kernel matrix row by row, using MCTS guided by a neural network policy. 
The algorithm starts with an all-zero kernel matrix. Then, at each step, the agent selects a column index where to place a one in the current row, aiming to satisfy the desired PDP constraints. 
Upon completion of a kernel candidate, the agent evaluates its decoding complexity under RMLD, and the result is used to shape the reward signal. 
Over time, the policy improves through self-play, progressively guiding the agent toward better-performing kernels.
%

\subsubsection{Training Loop}
\label{sec:Training Loop}
To train the neural network policy guiding the kernel construction, \algo{} follows a self-play training loop inspired by the AlphaZero framework. In each episode, the agent interacts with the kernel construction environment described in Algorithm~\ref{alg:AZenv}, attempting to build a valid $\ell \times \ell$ kernel $G$ that satisfies the target PDP \( \widetilde{\Dv}(\ell) \).
During the episode, the agent records the sequence of states, actions, and rewards encountered—this trajectory is referred to as \texttt{episodeData}.

After each episode, the resulting \texttt{episodeData} is appended to a cumulative training buffer. Periodically—every $K$ episodes—the accumulated data in this buffer is used to update the parameters of the policy and value networks, denoted by $f_\theta$, using standard supervised learning objectives \eqref{eqn:MCTS_loss}.
This procedure allows the agent to gradually refine its understanding of which actions lead to low-complexity kernels under the PDP constraints.

The training loop continues for a total of $N$ episodes. Over time, the agent improves its performance by reinforcing successful strategies and avoiding suboptimal choices, eventually converging to a policy capable of constructing high-performing kernels with significantly reduced decoding complexity. 


\begin{algorithm}
\caption{AlphaZero Self-Play Episode for Kernel Search}\label{alg:AZenv} 
\begin{algorithmic}[1]
\Function{kernelsearch}{$\widetilde{\Dv}(\ell), f_\theta$}
\State \textbf{Require:}  Target PDP $\widetilde{\Dv}(\ell)$, neural network $f_\theta$
\State \textbf{Ensure:}  List of (state, action, reward) tuples for training
\State $G = 0^{\ell \times \ell}$ \Comment{kernel state}
\State $k = 0$ \Comment{Current row index (bottom-up)}
\State $t = 0$ \Comment{Time step}
\State \texttt{episodeData} = [ ] \Comment{To store (state, action, reward)}
\While{$k < \ell$ and  $t < T$ }
    \State $t \gets t+1$ 
    \State $i = \ell-k-1$ \Comment{Current row index (top-down)}
    \State $r \gets -c$ \Comment{Step penalty}
    \State $j \gets \text{MCTS}(G, f_\theta)$ \Comment{Selection of column index using MCTS (Sec.~\ref{sec:Alpha Zero}).}
    \State $G[i, j] \gets 1$
    
    \If{$w_H(G[i]) == \widetilde{D}_i$} 
        \State $w \gets \min_{\mathbf{c} \in \langle G_{i+1}^{\ell-1} \rangle} d_H(G[i], \mathbf{c})$
        \If{$w == \widetilde{D}_i$}
            \State $r \gets \alpha$  \Comment{New row reward}
            \State $k \gets k + 1$ \Comment{Proceed to next row}
        \Else
            \State $G[i] \gets 0^\ell$ \Comment{Reset row}
        \EndIf
    \EndIf

    \If{$k == \ell$}
        \State $\Csf(G) \gets \text{RMLD}(G)$   \Comment{Calculate the RMLD decoding complexity}
        \State $r \gets$ \text{calc\_reward}($\Csf(G)$) 
    \EndIf
    \State \texttt{episodeData}.append(($G, j, r$))
\EndWhile


\State \Return \texttt{episodeData}
\EndFunction
\end{algorithmic}
\end{algorithm}



\subsubsection{Randomized initial steps} \label{sec:Rand_init_step}
To prevent the agent from repeatedly exploring identical kernels, we introduce randomness in the initialization phase by assigning ones to selected entries in the lower rows of the kernel $G$. Specifically, in \algoref{alg:AZenv}, after setting $G = 0^{\ell \times \ell}$, we randomly set a few bits in the bottom rows of $G$ to 1.   
For example, consider the case $\ell = 4$ and PDP $\widetilde{\Dv}(4) = [1, 2, 2, 4]$. Since the bottom row $G[3]$ must give $\widetilde{D}_3=4$, it is deterministically initialized as $G[3]=[1,1,1,1]$. The search then starts from the second-to-last row $G[2]$, corresponding to $\widetilde{D}_2=2$. To introduce diversity, we assign $G[2,j]=1$ where $j$ is chosen at random. For instance, if $j = 1$, the initial kernel could be:
$$
G = 
\left[\begin{array}{cccc}
0 & 0 & 0 & 0 \\
0 & 0 & 0 & 0 \\
0 & 1 & 0 & 0 \\
1 & 1 & 1 & 1 \\
\end{array}\right] 
\begin{array}{c}
  \\
  \\
\quad \leftarrow \text{current row } i=2  \\
  \\
\end{array}
$$

\subsection{Structure of the Rewards}   

The RL algorithm uses the following reward structure:
\begin{itemize}
  \item \textbf{Negative step reward -- $c$:} To encourage faster convergence, we penalize the RL agent for each step taken by assigning a negative reward of $-c$ per step, with $c > 0$. However, $c$ must be chosen carefully: if it is too large, the agent may prioritize minimizing the number of steps over reducing kernel complexity. In our experiments, we set $c = 0.1$. 
In each episode, if the episode length exceeds the maximum game length $T$, the search terminates. Without a negative step reward, the average search length tends to approach the maximum game length. Introducing a negative $c$ reduces the average episode length significantly, which in turn saves considerable training time. For example, with a kernel of size 16 and a maximum game length of 250, setting $c = 0$ results in an average episode length of about 240 steps. In contrast, setting $c$ to $0.1$ reduces the average length to around 60 steps per episode. Furthermore, the average episode length with $c = 0.1$ and $c = 1$ is nearly the same (around 60 steps), showing that a relatively small penalty (e.g., $c = 0.1$) is sufficient in our setting.

  
  \item \textbf{New row reward -- $\alpha$ :} Every time the RL agent finds a new row that satisfies the PDP constraint, it receives a reward $\alpha > 0$. If the agent successfully constructs all $\ell$ rows, the total reward from this term is $\ell \cdot \alpha$. We set $\alpha = 5$ for $\ell = 12$ and $\alpha = 10$ for $\ell = 16$.   
  
  \item \textbf{Complexity reward -- $r_\Csf$:} The reward based on kernel decoding complexity is computed using the transformation function: 
\begin{equation} \label{eq:comp_trans}
\begin{split}
  r_\Csf = \text{calc\_reward}(\Csf) = r_{\min} + (r_{\max} - r_{\min})  
  \cdot\left(\frac{\Csf_{\max} - \Csf}{\Csf_{\max}-\Csf_{\min}} \right)^{\gamma},
\end{split}
\end{equation}
where $r_{\min}$ and $r_{\max}$ denote the minimum and maximum reward, and $\Csf_{\min}$ and $\Csf_{\max}$ represent the observed bounds on kernel complexity. The value of $\Csf_{\max}$ is estimated using the random agent from \secref{sec:random_agent_search}. For instance, for $\ell=16$, we set $\Csf_{\max}=5000$, based on a maximum observed complexity of $5690$. The value of $\Csf_{\min}$ is determined from the lowest complexity kernel found so far; for $\ell=16$, we set $\Csf_{\min} = 1300$.   We set $r_{\min}=0$ and $r_{\max}=\Csf_{\max} - \Csf_{\min}$. 
The parameter $\gamma$ controls the nonlinearity of the reward: higher values emphasize low-complexity kernels. We use $\gamma = 2$.   
\end{itemize}

\medskip

By combining the rewards above, the total reward $v$ obtained upon reaching the top row of a kernel that satisfies the target PDP $\widetilde{\Dv}(\ell)$ is given by:  
\begin{equation} \label{eq:reward_shaping}
 v =- c \cdot (\text{steps}-\ell) + \text{calc\_reward}(\Csf) + \ell\cdot \alpha.
 \end{equation}

\subsection{Initializing the Bottom Rows with Pre-designed Sub-Kernels}
\label{sec:initialization}

As we shall discuss further in Sec.~\ref{sec:numerical_result}, the basic formulation of \algo{} proves effective for kernel sizes up to $\ell = 16$. However, scaling to larger kernel sizes remains challenging due to the exponential growth in the search space and the increased difficulty of satisfying the PDP constraints. 
To address this, we propose a general strategy that reduces decoding complexity and improves training efficiency by initializing the bottom rows of the kernel matrix with a fixed, low-complexity subkernel.

\vspace{0.2cm}
\noindent
The idea is to pre-fill the bottom $r$ rows of the $\ell \times \ell$ kernel with the rows of a known polarization kernel of size $r \times \ell$, denoted $K_r$. These rows are selected to satisfy the bottom portion of the relaxed PDP target \( \widetilde{\Dv}(\ell) \) and are not altered during training. The RL agent is then tasked only with constructing the top $\ell - r$ rows, reducing both the depth of the search tree and the variance of the training episodes. The resulting kernel has the form:
\[
G = \begin{bmatrix}
K_{\text{train}} \\
K_r
\end{bmatrix}, \quad \text{where } K_r \in \{0,1\}^{r \times \ell}.
\]
Here, $K_{\text{train}}$ denotes the trainable rows that are explored and optimized by \algo.

\vspace{0.2cm}
\noindent
This initialization scheme brings several advantages:
\begin{itemize}[leftmargin=*]
    \item It significantly reduces the effective search depth from $\ell$ to $\ell - r$, making the learning task easier for the agent;
    \item The fixed subkernel $K_r$ can be chosen to contribute low-complexity behavior and favorable polarization properties to the overall kernel; 
    \item It allows modular kernel construction by building larger kernels atop smaller, well-understood ones.
\end{itemize}

\vspace{0.2cm}
\noindent
An example of this approach is detailed in Sec.~\ref{sec:numerical_result}, where the bottom five rows of a size-16 kernel are initialized using the sorted Arıkan kernel $K_5$. Empirical results show that this strategy leads to significantly lower decoding complexity compared to training a full size-16 kernel from scratch.

\vspace{0.2cm}
\noindent
While the proposed scheme is most naturally applicable to kernel sizes $\ell$ that are multiples or powers of 2—where existing low-complexity kernels such as $K_4$, $K_5$, or $K_8$ are available—the extension to arbitrary sizes remains an open research direction.

\begin{example}
For the size 16 kernel, we can further reduce the complexity by initializing the 5 bottom rows with sorted Arıkan’s kernel $K_5$. That is, $K_5 = S_{16}[11:15]$.  
\vspace{0.15cm}   

In \tabref{tab:rand_comp_hand}, the random search results for a size-16 kernel with $r$ bottom rows initialized to $K_5$ are reported. $r=1$ corresponds to no initialization, since the last row must be an all-ones vector. The kernel initialized with 5 bottom rows ($r=5$) exhibits lower complexity compared to the one without initialization ($r=1$).

However, this method is applicable only to kernels of size \(2^N\), where \(N \in \mathbb{N}_{>0}\). The structure of low-complexity bottom rows for kernels of other sizes (e.g., size \(12\)) remains an open problem. 
\end{example}

\subsection{Training Different Kernel Sizes in a Single Run}
\label{sec:multi_kern_single_run}
As demonstrated in \cite{AlphaTensor}, kernels of different sizes can be trained in a single run. To achieve this, we first select a set of target kernel sizes (e.g., from size 4 to size 16). At the start of each game, we randomly sample a kernel size from this set and optimize the corresponding kernel.

To ensure consistent input dimensions for the neural network, we resize smaller kernels to match the maximum kernel size (e.g., 16) by padding unused rows and columns with zeros.


Training kernels of different sizes in a single run improves the efficiency of the training compared to the original setting, in which the kernels are trained individually. However, although \cite{AlphaTensor} reports that this scheme outperforms the original setting, we observed cases where the original setting performs better (e.g., the size-16 kernel). Moreover, the agent can still become trapped in a suboptimal kernel, possibly due to unbalanced hyperparameters and reward shaping across different kernel sizes. The selection of appropriate hyperparameters remains the primary challenge for this scheme.

\begin{table}[h]
\captionsetup{justification=centering}
\caption{
Minimum and maximum RMLD complexities of a size-16 kernel satisfying the target PDP 
$\widetilde{\Dv}(16)$,\\
found by a random agent with $r$ bottom rows initialized from the sorted Arıkan kernel $K_5$. 
}
\begin{center}
\begin{tabular}{|c|c|c|c|}
\hline
$r$ & Min & Max & Iterations\\ 
\hline
%
1  & 3338 & 7886 & 10K\\
2  & 1636 & 6162 & 10K\\
3  & 1592 & 6182 & 10K\\
4  & 1452 & 6182 & 10K\\
5  & 1516 & 6182 & 10K\\ 
1  & 2120 & 7906 & 800K\\
4  & 1388 & 6468 & 800K\\
5  & 1360 & 6468 & 800K\\
\hline
\end{tabular}
\label{tab:rand_comp_hand}
\end{center}
\vspace{-0.25cm}
\end{table}
\vspace{0.15cm}

\section{Baselines}
\label{sec:Baselines}

To benchmark the performance of the proposed reinforcement learning framework \algo, we first consider two baseline methods for solving the kernel design problem formulated in~\eqref{eq:K minimize C}. Specifically, both methods aim to identify feasible polarization kernels—i.e., binary matrices \( G \in \{0,1\}^{\ell \times \ell} \) whose PDP satisfies the relaxed constraint \( \widetilde{\Dv}(\ell) \) provided in Table~\ref{tab:pdp}.

Due to the discrete nature of the PDP, it is non-trivial to define a continuous loss or distance function over candidate kernels. As a result, we adopt two complementary strategies for exploring the kernel space:
\begin{itemize}[leftmargin=*]
    \item a deterministic, exhaustive \emph{brute-force search};
    \item a stochastic \emph{random sampling agent}.
\end{itemize}
These two approaches are outlined below.

\subsection{Brute-Force Kernel Search}
\label{sec:brute_force}

We first consider a brute-force search algorithm adapted from~\cite{TroBrute24}. The goal is to construct a kernel row by row such that each row satisfies its corresponding entry in the target PDP.
%
%
The kernel is constructed row by row, starting from the bottom row \( i = \ell-1 \) and moving upward.

Let \( \mathcal{M}_w \) denote the set of all binary vectors of Hamming weight \( w \), ordered lexicographically. The algorithm proceeds in reverse, starting from the bottom row \( i = \ell - 1 \), and at each step attempts to construct a vector \( v_i \in \mathcal{M}_{\widetilde{D}_i} \) such that:
\begin{equation}
\min_{v \in \langle v_{i+1}, \dots, v_{\ell-1} \rangle} d_H(v_i, v) = \widetilde{D}_i. \label{eq:bf_condition}
\end{equation}
That is, the minimum Hamming distance between \( v_i \) and the subcode generated by the rows $i+1,\dots, \ell-1$ must match the target PDP value \( \widetilde{D}_i \).
%
%
If such a row \( v_i \) exists, it is accepted and the algorithm proceeds to the next row \( i - 1 \). Otherwise, it backtracks to row \( i + 1 \) and tries the next candidate in \( \mathcal{M}_{\widetilde{D}_{i+1}} \). 
Ties or multiple valid options can be resolved by fixed ordering (e.g., lexicographic) or random selection.
This depth-first backtracking ensures completeness: either a feasible kernel is found, or all configurations are exhausted under a predefined search budget.

\subsection{Random Agent Kernel Search}
\label{sec:random_agent_search}

We next consider a stochastic alternative to brute-force construction. Here, a randomized agent builds the kernel row by row by flipping randomly selected bits, subject to the PDP constraints.
This approach arches back to the random agent discussed in Sec. \ref{sec:Random Agent}.
The agent starts with a zero matrix \( G \in \{0,1\}^{\ell \times \ell} \) and proceeds bottom-up (i.e., starting from row \( i = \ell - 1 \)). At each step:
\begin{itemize}[leftmargin=*]
    \item the agent samples one or more bit positions in row \( i \) to set to 1;
    \item if the resulting row has Hamming weight \( \widetilde{D}_i \), the agent checks whether it satisfies the minimum distance condition in~\eqref{eq:bf_condition};
    \item if so, the row is accepted and the process moves to row \( i - 1 \); otherwise, the row is reset and re-sampled.
\end{itemize}

This randomized construction continues until either a valid kernel is found or a maximum number of trials is reached. 
Unlike brute-force, this method does not guarantee completeness but provides a flexible tool that provides insights on the complexity of the minimization problem in \eqref{eq:K minimize C}.
In particular, the random agent effectively  samples the distribution of feasible kernels which provide the opportunity of an empirical analysis of the RMLD complexity over the feasible set.
This also results in an estimate of the lower and upper bounds on RMLD decoding complexity across valid kernels sizes. 

%


\section{Numerical Results} 
\label{sec:numerical_result}

In this section, we present numerical results obtained with \algo, and compare them with the baselines provided by  the brute force agent in \secref{sec:brute_force} and the random agent introduced in \secref{sec:random_agent_search}.

\subsection{Empirical Analysis}

As hinted in \secref{sec:random_agent_search}, we can use the random agent as a way to sample the feasible set of kernels which satisfy the PDP condition and observe the distribution of the complexity over this set.  
Table~\ref{tab:rand_comp} reports the minimum and maximum decoding complexities of kernels of size $\ell$ obtained using the random agent search.
\begin{table}[h]
\caption{Minimum and maximum RMLD complexity found by the random agent in \secref{sec:random_agent_search}.}
\begin{center}
\begin{tabular}{|c|c|c|c||c|c|c|c|}
\hline
\textbf{$\ell$} & Min & Max & Iterations & \textbf{$\ell$} 
& Min & Max & Iterations \\ 
\hline
%
4  & 32   & 44   & 10K   & 12 & 888  & 1834   & 10K \\
5  & 57   & 93   & 10K   & 12  & 784  & 1834 & 200K \\
6  & 88   & 158  & 10K   & 13  & 1137 & 2731 & 10K \\
7  & 121  & 249  & 10K   & 14  & 1686 & 3758 & 10K \\
8  & 156  & 340  & 10K   & 15  & 2245 & 5669 & 10K \\
9  & 291  & 603  & 10K   & 16  & 3338 & 7886 & 10K \\
10 & 356  & 854  & 10K   & 16  & 2120 & 7906 & 400K \\
11 & 607  & 1311 & 10K   & 16  & 2120 & 7906 & 800K \\
\hline
\end{tabular}
\label{tab:rand_comp}
\end{center}
\vspace{-0.25cm}
\end{table}

\begin{figure}[h!]
\centering
\resizebox{0.5\textwidth}{!}{\input{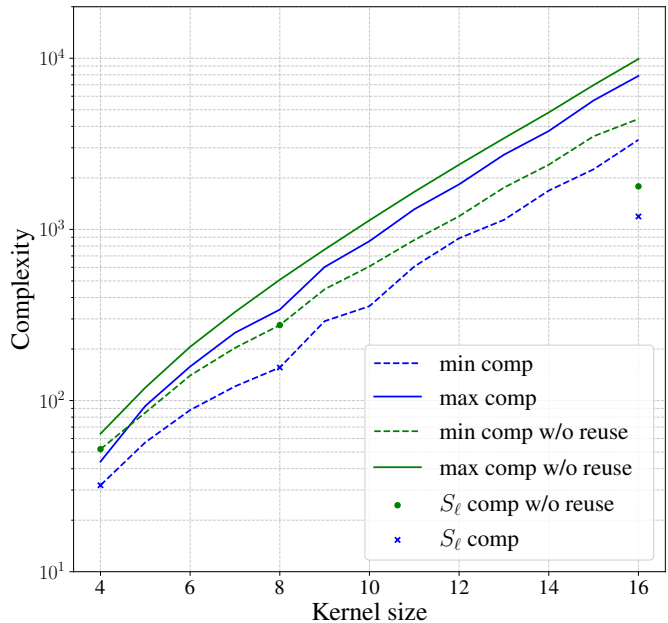}}
\caption{
Complexity of kernels of different sizes. The blue and green lines represent the minimum and maximum complexities obtained from the random agent with $10{,}000$ Monte Carlo iterations. The blue line corresponds to kernels decoded using the trellis-reuse method, as described in \secref{sec:trellis_reuse}, while the green line corresponds to kernels decoded without trellis reuse in the RTPA decoding process. Additionally, the blue and green dots indicate the decoding complexities of the sorted Arıkan kernels.
}
\label{fig:min_max_comp}
\end{figure}

In Fig.~\ref{fig:min_max_comp} we display in blue color the data of Table~\ref{tab:rand_comp}. 
A few observations are in order.
For kernel size $\ell=12$ and $\ell=16$, we repeat the random agent simulation with different number of iterations: $[10K,\,200K]$ and $[10K,\,400K,\,800K]$, respectively. 
Clearly, as the kernel size increases, the random agent's ability to explore the space of solutions decreases exponentially. 
For this reason, one cannot rely on this naive approach for solving the optimization problem at hand. 
While we are unable to prove this fact, it appears that the maximal and minimal complexity have a logarithmic scaling with the kernel size, as can be seen from Fig. \ref{fig:min_max_comp}. 
Using a simple linear approximation, we obtain the bounds of the form 
\ea{
\Csf_i \approx 2^{\al \ell + \beta},
}
where
\eas{
\Csf_{\max} & = 2^{0.600 \ell + 3.581},\\
\Csf_{\min} & = 2^{0.547 \ell + 3.062}.
}
%
%
%



Another use we can make of the random kernel sampling offered by the random agent is that of comparing the reduction in complexity offered by the trellis reuse of RMLD. 
This is depicted in Fig. \ref{fig:min_max_comp}: here we plot the maximal and minimal complexity with trellis reuse--as in Table \ref{tab:rand_comp}-- together with the same maximal and minimal complexity without trellis reuse. 
The blue and green lines represent the minimum and maximum complexities obtained with the random agent, where each kernel size $\ell$ is assigned a given PDP $\widetilde{\Dv}(\ell)$ as shown in Table~\ref{tab:pdp}. We fix the number of iterations to 10K for all values of $\ell$. 
%
%
It is rather clear that the trellis reuse offers an exponential decrease in complexity for a randomly selected kernel as the kernel size increases.
In the figure, we also plot the complexity of the sorted Arıkan’s kernels-- which have error exponent equal to 0.5-- for $\ell = 4, \ell=8$ and $\ell=16$. 
We plot the complexity of these kernels with and without applying the trellis reuse method, as described in \secref{sec:trellis_reuse}.
%
Note that the minimum complexity at kernel sizes 4 and 8 matches those of $S_4$ and $S_8$, implying that the random agent is able to recover low-complexity kernels when the kernel size is relatively small.
For $\ell = 16$, $S_{16}$ achieves a much lower complexity than the minimum complexity obtained from the random agent, at the cost of a lower error exponent (0.5) compared to the error exponent $E(\widetilde{\Dv}(16)) = 0.5183$ of the kernel found by the random agent for the same kernel size.






Before moving forward, we compare the empirical distribution complexity found by the random agent with that obtained by using the 
kernel initialization discussed in  \secref{sec:initialization}.
Our goal here is in arguing that the kernel initialization technique is effective in eliminating the bulk of the kernels with high complexity, while not substantially affecting the long tail of the distribution toward the low complexity kernels. 

\begin{figure}
    \centering
    \resizebox{0.5\textwidth}{!}{\input{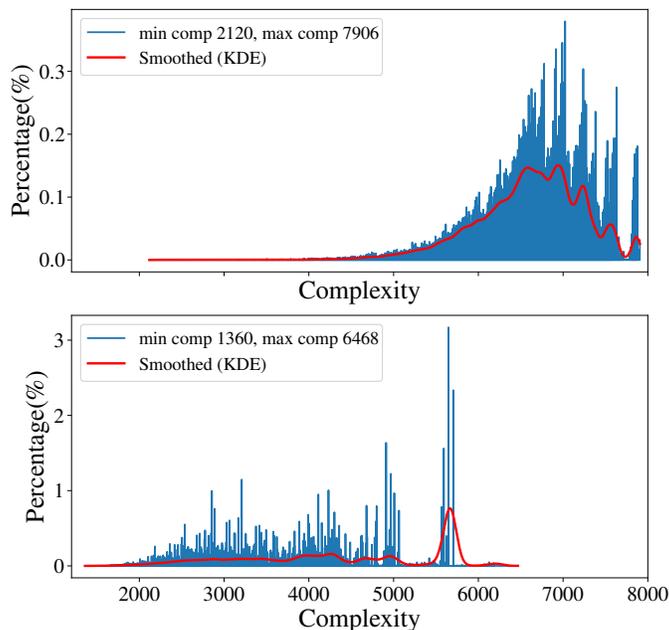}}
    \vspace{-0.35cm}
    \caption{Complexity spectrum of random agent search. The red line represents 
    a smoothed version of the complexity spectrum. The upper plot corresponds to the setting 
    without bottom-row initialization, while the lower plot corresponds to the setting where the 
    bottom five rows are initialized with the sorted Arıkan kernel $K_5$.}
    \label{fig_comp_spec_G16}
    \vspace{-0.1cm}
\end{figure}

The complexity spectrum of the size-16 kernels obtained by the random agent is shown in the upper plot of Fig.~\ref{fig_comp_spec_G16}, with RMLD complexities ranging from $2120$ to $7906$. By initializing the bottom $5$ rows with $K_5$, the spectrum improves, as illustrated in the lower plot of Fig.~\ref{fig_comp_spec_G16}, with RMLD complexities ranging from $1360$ to $6468$.


\subsection{\algo{} Agent }
\begin{figure}
    \centering
    \resizebox{0.48\textwidth}{!}{\input{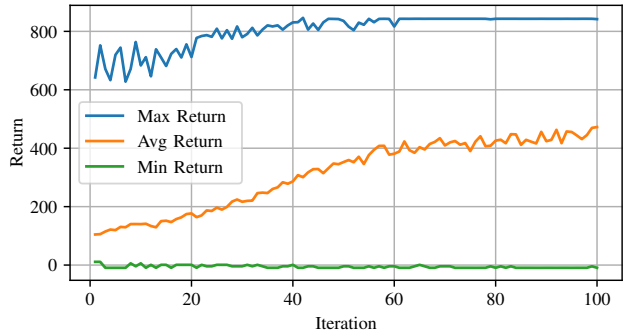}}
    \vspace{-0.35cm}
    \caption{Minimum, maximum, and average total rewards during training for $\ell=12$. Each iteration consists of $2000$ self-play games, with a total of  $2\cdot 10^5$ episodes over $100$ iterations.}
    \label{fig_12}
    \vspace{-0.1cm}
\end{figure}

We now focus on the results obtained using the \algo{} agent for two kernel sizes: $\ell = 12$ and $\ell = 16$.\footnote{The code-base used for the experiments in this section is publicly available to support reproducibility: \url{https://github.com/jaco267/AlphaPolar}.
}
\figref{fig_16} shows the evolution of total rewards during training for $\ell = 16$ and with PDP  $\widetilde{\Dv}(16)$ from \tabref{tab:pdp}. At each iteration, $K = 2000$ self-play games are generated  as in Sec. \ref{sec:Training Loop}, and the neural network $f_\theta$ is trained.
The average reward stabilizes after approximately 230 iterations, suggesting convergence. For smaller values of $\ell$, convergence occurs faster; 
for example, for $\ell = 12$ in \figref{fig_12}, convergence is reached after approximately 100 iterations.

The best \algo-found kernel for $\ell = 16$ without bottom-row initialization has decoding complexity 1624, reported as $A_{16}$ in Table \ref{tab:alpha} in Appendix \ref{app:Tabulated Large Polarization Kernels}.
$A_{16}$ was discovered using the approach in Sec. \ref{sec:Rand_init_step}
 and starting from the initial kernel $A_{16}^{[\mathsf{Init}]}$ as tabulated in Table \ref{tab:init} in Appendix \ref{app:Tabulated Large Polarization Kernels}. 

\begin{figure}
    \centering
    \resizebox{0.48\textwidth}{!}{\input{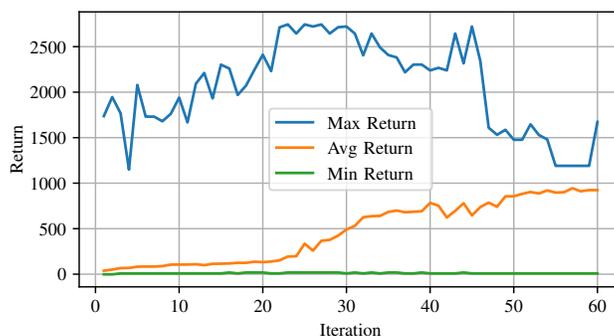}}
    \vspace{-0.5cm}
    \caption{Minimum, maximum, and average total rewards during training for $\ell=16$. With bottom rows initialized to sorted Arıkan’s kernel $K_5$.  Each iteration consists of 2000 self-play games, with a total of $1.2 \cdot 10^5$ episodes over 60 iterations.}
    \label{fig_16h}
    \vspace{-0.5cm}
\end{figure}

We further optimize the size-16 kernel by initializing its bottom rows with the sorted Arıkan kernel $K_5$, following the approach described in \secref{sec:initialization}.
During training, we randomly select the number of bottom rows to be 3, 4, or 5 with equal probability. The training curve in this setting is shown in Fig. \ref{fig_16h}. It can be observed that the maximum return reaches its peak at around 30 iterations. The minimum-complexity kernel $A_{16h}$ was obtained at iteration 27, which is close to the point of maximum return in this run. 
%
%
The resulting kernel $A_{16h}$ is listed in Table~\ref{tab:init} in Appendix~\ref{app:Tabulated Large Polarization Kernels}.
The four bottom rows of $A_{16h}$ are identical to those of $K_5$.  
The kernel $A_{16h}$ achieves a decoding complexity of 1308, representing approximately a $17\%$ improvement compared to the handcrafted kernel $H_{16}$ from \cite{handcraft_win} (see Table~\ref{tab:hand} in Appendix~\ref{app:Tabulated Large Polarization Kernels}), which achieves complexity 1580 under our RMLD complexity evaluation. It is important to note that  kernel $H_{16}$ was not designed for the RMLD algorithm. The reported complexity of 1580 corresponds to the decoding complexity obtained when applying the RMLD algorithm to the handcrafted kernel, as described in \secref{sec:trellis_reuse}.

For $\ell=12$, the best decoding complexity observed is 764, which corresponds to  kernel  $A_{12}$ in Table \ref{tab:alpha} in Appendix \ref{app:Tabulated Large Polarization Kernels}.  
Similarly to $A_{16}$, $A_{12}$ was found starting from the initialized kernel $A_{12}^{[\mathsf{Init}]}$ in Table \ref{tab:init} in Appendix \ref{app:Tabulated Large Polarization Kernels}.


\begin{figure}
    \centering
    \resizebox{0.48\textwidth}{!}{\input{images/returns_plot_16_v2.pgf}}
    \vspace{-0.5cm}
    \caption{Minimum, maximum, and average total rewards during training for $\ell=16$. Each iteration consists of 2000 self-play games, with a total of $5 \cdot 10^5$ episodes over 250 iterations.}
    \label{fig_16}
    \vspace{-0.5cm}
\end{figure}

\begin{figure}
    \centering
    \resizebox{0.48\textwidth}{!}{\input{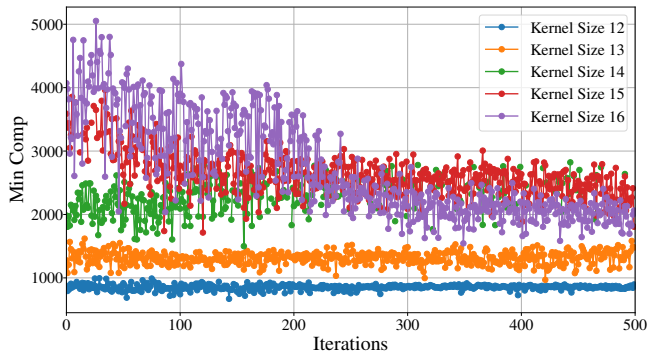}}
    \vspace{-0cm}
    \caption{Training multiple kernels in a single run. The Y-axis is the minimum complexity for size 12 to 16 kernels in the corresponding iteration. Each iteration have 2000 episodes. 
    }
    \label{fig_multi_kern_comp}
    \vspace{-0cm}
\end{figure}

%



\begin{table}[tbp]
\captionsetup{justification=centering}
\caption{Minimum kernel complexity found by \algo.\\
Numbers in boldface indicate kernels with complexity lower than those in \tabref{tab:rand_comp}.} 
\begin{center}
\begin{tabular}{|c|c|c|c|c|c|c|c|c|c|c|c|c|c|}
\hline
\textbf{$\ell$} & 4 & 5 & 6 & 7
& 8 & 9 & 10 & 11 & 12 & 13 &14&15&16h\\ 
\hline
 Complexity   & 32   & 57   & 88   & 121 & 156 & $\mathbf{277}$ & $\mathbf{316}$ & $\mathbf{509}$ & $\mathbf{764}$ & 1157 & $\mathbf{1548}$ & 2331 & $\mathbf{1308}$ \\
 \hline
  \( E(\widetilde{\Dv}) \) & 0.5 & 0.43   & 0.45 & 0.45 & 0.5 & 0.46 & 0.47 & 0.48 & 0.48 & 0.48 & 0.49& 0.49 & 0.5183 \\
\hline
\end{tabular}
\label{tab:min_kern_az}
\end{center}
\vspace{-0.25cm}
\end{table}

The training time of \algo \space for the $A_{16}$ kernel is approximately 47 hours for 150 iterations and about 105 hours for 250 iterations. In contrast, Random Search without handcrafted kernel initialization for a size-16 kernel requires roughly 112 hours to complete 800K iterations, resulting in a minimum complexity of 2120, which is higher than the complexity of $A_{16}$ (1624). Hence, under comparable search times, \algo \space is able to achieve kernels with lower complexity than Random Search.

\figref{fig_multi_kern_comp} illustrates the minimum-complexity kernels obtained by \algo\ at various training iterations using the multi-kernel training procedure described in Sec.~\ref{sec:multi_kern_single_run}. The agent is initialized with kernel sizes drawn from the following probability distribution: size-16 (0.40), size-15 (0.29), size-14 (0.15), size-13 (0.11), and size-12 (0.05). Smaller kernels are assigned lower probabilities since they generally require fewer iterations to converge.


Table~\ref{tab:min_kern_az} reports the minimum-complexity kernels discovered by \algo. Here, “16h” denotes the complexity of the $A_{16h}$ kernel, while “12” refers to the complexity of the $A_{12}$ kernel. 
The remaining kernel sizes were trained jointly in a single experiment, consistent with the multi-kernel training setup in Sec.~\ref{sec:multi_kern_single_run}. The kernels correspond to Table~\ref{tab:min_kern_az} are denoted as $A_9, A_{10}, A_{11}, A_{12}, A_{14}$, and $ A_{16_h}$ in Tables \ref{tab:alpha} and  \ref{tab:init} in Appendix \ref{app:Tabulated Large Polarization Kernels}.

\subsection{Block Error Rate performance}
We compare the block error rate (BLER) performance of $(n=256, k=128)$ codes constructed using: (i) Arıkan’s $\ell=2$ kernel; (ii) the handcrafted kernel $H_{16}$ from \cite{handcraft_win}; and (iii) the \algo-found kernel $A_{16}$. The frozen bit sets are optimized at each SNR. Arıkan’s code is decoded using SCD, while $H_{16}$ and $A_{16}$ are decoded using RMLD.

\begin{figure}
    \centering
    \begin{tikzpicture}
\begin{axis}[
    xlabel={$E_b/N_0$ [dB]},
    ylabel={BLER},
    ymode=log,
    xmin=0, xmax=4.5,
    ymin=1e-6, ymax=1,
    legend pos=south west,
    grid=both,
    width=8cm,
    height=6cm
]

\addplot+[mark=*] coordinates {
    (0.0, 8.4202e-01)  
    (0.5, 6.6234e-01)
    (1.0, 4.1924e-01)
    (1.5, 2.1348e-01)
    (2.0, 7.9260e-02)  
    (2.5, 2.2200e-02)
    (3.0, 3.9400e-03)
    (3.5, 4.4300e-04)
    (4.0, 4.4333e-05)  
    (4.5, 1.5000e-06)     
};
\addlegendentry{RMLD $H_{16}$}

\addplot+[mark=triangle*] coordinates {
    (0.0, 8.2158e-01)    
    (0.5, 6.2598e-01)
    (1.0, 3.8660e-01)
    (1.5, 1.8264e-01)
    (2.0, 6.1500e-02)    
    (2.5, 1.5470e-02)
    (3.0, 2.7500e-03)
    (3.5, 3.6357e-04)
    (4.0, 3.3333e-05)
    (4.5, 1.0000e-06)
};
\addlegendentry{RMLD $A_{16}$}

\addplot+[mark=square*] coordinates {
    (0.0, 8.8312e-01)  
    (0.5, 7.3074e-01)
    (1.0, 5.1536e-01)
    (1.5, 2.9734e-01)
    (2.0, 1.3056e-01)  
    (2.5, 4.6900e-02)
    (3.0, 1.2060e-02)
    (3.5, 2.4289e-03)
    (4.0, 4.0898e-04) 
    (4.5, 4.3333e-05)
};
\addlegendentry{Arıkan's $2\times 2$ kernel}

\end{axis}
\end{tikzpicture}
    \vspace{-0.25cm}
    \caption{Block Error Rate (BLER) performance of polar codes using Arıkan’s kernel (SC decoding), the handcrafted kernel $H_{16}$ from \cite{handcraft_win}, and the \algo-found kernel $A_{16}$. At each SNR, frozen bits are selected based on $3\times 10^6$ MC simulations. BLER curves are estimated using $3\times 10^6$ MC iterations per SNR.}
    \label{fig_perf_A16}
    \vspace{-0.5cm}
\end{figure}
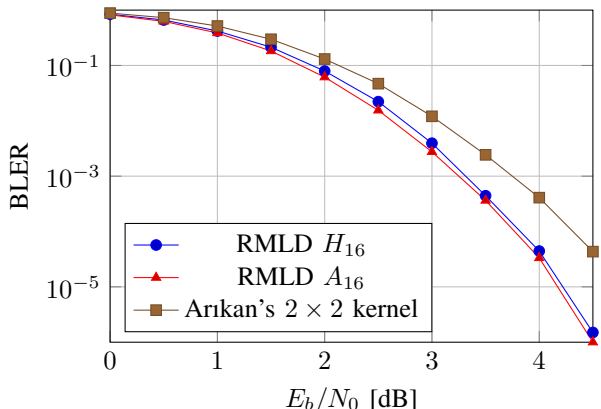

As shown in \figref{fig_perf_A16}, the kernels $H_{16}$ and $A_{16}$ achieve similar BLER performance, both outperforming Arıkan’s code. This improvement is consistent with their superior error exponents, $E_{H_{16}} = E_{A_{16}} \approx 0.5183$, compared to $E_{\text{Arıkan}} = 0.5$.

\section{Conclusion and Open Problems}
\label{sec:conclusion}
   
In this work, we have proposed a search method for large polarization kernels, termed \algo.
\algo\ is based on the AlphaZero algorithm. Given a target PDP, \algo\ is able to discover kernels with lower RMLD complexity compared to the random search method.  We have also introduced several techniques to improve the robustness of the search process, including randomized initial steps and initializing the bottom rows with pre-designed sub-kernels.         

However, several open problems remain regarding the search method for large polarization kernels:  
\begin{itemize}[leftmargin=*,noitemsep]
    \item \emph{ Non-monotonic PDP:} 
In our setting, we have restricted our attention to monotonic-increasing PDPs. Nevertheless, it is possible to achieve lower complexity while preserving the same error exponent by employing non-monotonic PDPs. 
However, the search space for non-monotonic PDPs is vast, making the problem highly challenging. The systematic exploration of low-complexity non-monotonic PDPs remains an open problem.
    \item \emph{ Other design metrics:} 
    In this work, we have focused primarily on the error exponent and decoding complexity. Other design metrics, such as the scaling exponent or memory complexity, also merit further investigation.
    \item \emph{ Kernel with Larger size:} 
    In our experiments, we have restricted the kernel size to \( \ell \leq 16 \). 
When searching for larger kernels (e.g., of size 17), the number of valid kernels that satisfy the PDP constraint becomes sparse, 
and the data collection process in \algo\ requires significantly longer training time. 
In such cases, a naive depth-first search becomes inefficient.
Several optimization techniques proposed in~\cite{TroBrute24} can accelerate the search process, 
but they are generally effective only for kernel sizes up to approximately 29. 
For even larger kernels (e.g., \( \ell = 64 \)), since the episode length scales with \( \ell^2 \), 
the training time becomes prohibitively long. 
In these cases, the generalized concatenation method described in~\cite{tri64} may be required to effectively search for kernels of larger sizes.
\end{itemize}

\ifCLASSOPTIONcaptionsoff
  \newpage
\fi

\ifCLASSOPTIONcaptionsoff
  \newpage
\fi


\appendices

\section{Tabulated Large Polarization Kernels}
\label{app:Tabulated Large Polarization Kernels}

In the following, we tabulate the relevant matrices discussed in the paper and in the related literature. 
More precisely:
\begin{itemize}
    \item Table \ref{tab:sorted} -- {\textbf{Sorted Arıkan's kernels:}} For \( i \in \{4, 8, 16\} \), \( S_i \) denotes the sorted version of Arıkan’s kernel in \cite{TriRTPA23}. 
    \item Table \ref{tab:hand} --
    {\textbf{Handcrafted kernels:}}  $H_{16}$  represents the sorted version of the handcrafted kernel presented in \cite{handcraft_win}. Certain rows have been permuted to reduce the RMLD decoding complexity.
    \item  Table \ref{tab:alpha} -- {\textbf{\algo \space kernels:}} for \( i \in \{9, 10, 11, 12, 14, 16\} \), \( A_i \) refers to the kernels discovered using \algo. 
    \item Table \ref{tab:init} --  \textbf{Initialization sub-matrices} For $i$ in \{12,16\}. This refers to the submatrices used for \algo's initialization as discussed in Sec.  \ref{sec:initialization}.
    % ...
\end{itemize}
\begin{table*}[!htbp]
\caption{Sorted Arıkan's kernels}
\label{tab:sorted}
\centering
\resizebox{0.65\textwidth}{!}{%
\begin{tabular}{lll}
\toprule
\Sfour & \multicolumn{2}{l}{\Seight} \\
& \\
\multicolumn{3}{l}{\Ssixteen}    \\
& \\
\bottomrule
\end{tabular}
}
\end{table*}

\begin{table*}[!htbp]
\caption{Handcrafted kernel \cite{handcraft_win}}
\label{tab:hand}
\centering
\resizebox{0.75\textwidth}{!}{%
\begin{tabular}{lll}
\toprule
\multicolumn{3}{l}{\Hsixteen}    \\
& \\
\bottomrule
\end{tabular}
}
\end{table*}

\begin{table*}
\caption{ \algo's kernels}
\label{tab:alpha}
\centering
\resizebox{0.85\textwidth}{!}{%
\begin{tabular}{llll}
\toprule
\multicolumn{1}{l}{\Anine}  & \multicolumn{3}{l}{\Aten} \\
\\
\multicolumn{1}{l}{\Aeleven} & \multicolumn{3}{l}{\Afourteen}    \\
\\
\Atwelve & \Asixteen &    \\
\bottomrule
\end{tabular}
}
\end{table*}

\begin{table*}
\caption{ \algo's initialization sub-kernels}
\label{tab:init} 
\centering
\resizebox{0.75\textwidth}{!}{%
\begin{tabular}{l}
\toprule
\AtwelveInit\\
\\
\AsixteenInit \\
\\
\AsixteenHand \\
\\
\bottomrule
\end{tabular}
}
\end{table*}

\section{Size-4 Extended  Kernel Construction}
\label{app:size 4}
In this appendix we provide an example of the extended kernel construction in Sec. \ref{sec: Extended Kernel Construction} for a kernel of size 4, $S_4$ in Table  \ref{tab:sorted} in Appendix \ref{app:Tabulated Large Polarization Kernels}. 


The extended matrices of $S_4$ for each decoding phase are:
\setcounter{paragraph}{0}
\paragraph{Phase 0}
\[
S_4^{(0)} =
\begin{bmatrix}
1 & 0 & 0 & 0 & \vrule & 1 \\
1 & 0 & 1 & 0 & \vrule & 0 \\
1 & 1 & 0 & 0 & \vrule & 0 \\
1 & 1 & 1 & 1 & \vrule & 0
\end{bmatrix}
\]

\paragraph{Phase 1}
\[
S_4^{(1)} =
\begin{bmatrix}
1 & 0 & 1 & 0 & \vrule & 1 \\
1 & 1 & 0 & 0 & \vrule & 0 \\
1 & 1 & 1 & 1 & \vrule & 0
\end{bmatrix}
\]

\paragraph{Phase 2}
\[
S_4^{(2)} =
\begin{bmatrix}
1 & 1 & 0 & 0 & \vrule & 1 \\
1 & 1 & 1 & 1 & \vrule & 0
\end{bmatrix}
\]

\paragraph{Phase 3}
\[
S_4^{(3)} =
\begin{bmatrix}
1 & 1 & 1 & 1 & \vrule & 1
\end{bmatrix}
\]
For a kernel $G_\ell$, its extended matrices $G_\ell^{(i)}$ define the valid codewords for evaluating the maximum in \eqref{eq:sc_decoding} over the subset of codewords consistent with a given value of \( u_i \), where $G^{(i)}_\ell$ is used to decode $u_i$. 


\clearpage
\section{Derivation of LLR Domain Approximation for MAP expression}
\label{app:MAP_der}
We can write down the LR domain version of \eqref{eq:sc_decoding} as 
$$
\hat{l}_i = \text{exp}(\hat{L}_i) = \frac{\sum_{\bar{c}^{(i)} \in \langle G_\ell^{(i)} \rangle\,:\, \bar{c}^{(i)}_\ell = 0} \prod_{j=0}^{\ell-1} \Pr(c_j = \bar{c}^{(i)}_j \mid y_j)}{\sum_{\bar{c}^{(i)} \in \langle G_\ell^{(i)} \rangle\,:\, \bar{c}^{(i)}_\ell = 1} \prod_{j=0}^{\ell-1} \Pr(c_j = \bar{c}^{(i)}_j \mid y_j)} \\
= \frac{\sum_{\bar{c}^{(i)} \in \langle G_\ell^{(i)} \rangle\,:\, \bar{c}^{(i)}_\ell = 0} \prod_{j=0}^{\ell-1} \Pr(c_j = \bar{c}^{(i)}_j \mid y_j)}{\sum_{\bar{c}^{(i)} \in \langle G_\ell^{(i)} \rangle\,:\, \bar{c}^{(i)}_\ell = 1} \prod_{j=0}^{\ell-1} \Pr(c_j = \bar{c}^{(i)}_j \mid y_j)} \frac{\frac{1}{\prod_{j=0}^{\ell-1} \Pr(c_j = 1 \mid y_j)}} {\frac{1}{\prod_{j=0}^{\ell-1} \Pr(c_j = 1 \mid y_j)}}\\
$$
$$
= \frac{\sum_{\bar{c}^{(i)} \in \langle G_\ell^{(i)} \rangle\,:\, \bar{c}^{(i)}_\ell = 0} \prod_{j=0}^{\ell-1}f(l_j|\bar{c}_j^{(i)})}{\sum_{\bar{c}^{(i)} \in \langle G_\ell^{(i)} \rangle\,:\, \bar{c}^{(i)}_\ell = 1} \prod_{j=0}^{\ell-1} f(l_j|\bar{c}_j^{(i)})} ,  f(l_j \mid \bar{c}_j^{(i)}) =
\begin{cases}
l_j, & \text{if } \bar{c}_j^{(i)} = 0 \\
1,   & \text{if } \bar{c}_j^{(i)} = 1
\end{cases}
$$
$$
= \frac{\sum_{\bar{c}^{(i)} \in \langle G_\ell^{(i)} \rangle\,:\, \bar{c}^{(i)}_\ell = 0} \prod_{j=0}^{\ell-1}\text{exp}(F(L_j|\bar{c}_j^{(i)}))}{\sum_{\bar{c}^{(i)} \in \langle G_\ell^{(i)} \rangle\,:\, \bar{c}^{(i)}_\ell = 1} \prod_{j=0}^{\ell-1} \text{exp}(F(L_j|\bar{c}_j^{(i)}))} = \frac{\sum_{\bar{c}^{(i)} \in \langle G_\ell^{(i)} \rangle\,:\, \bar{c}^{(i)}_\ell = 0} \text{exp}(\sum_{j=0}^{\ell-1}F(L_j|\bar{c}_j^{(i)}))}{\sum_{\bar{c}^{(i)} \in \langle G_\ell^{(i)} \rangle\,:\, \bar{c}^{(i)}_\ell = 1} \text{exp}(\sum_{j=0}^{\ell-1} F(L_j|\bar{c}_j^{(i)}))}
$$
$$
\approx \frac{\max_{\bar{c}^{(i)} \in \langle G_\ell^{(i)} \rangle \,:\, \bar{c}^{(i)}_\ell = 0} \text{exp}(\sum_{j=0}^{\ell-1}F(L_j|\bar{c}_j^{(i)}))}{\max_{\bar{c}^{(i)} \in \langle G_\ell^{(i)} \rangle\,:\, \bar{c}^{(i)}_\ell = 1} \text{exp}(\sum_{j=0}^{\ell-1} F(L_j|\bar{c}_j^{(i)}))}
$$
After take the logarithm on both sides, we obtain \eqref{eq:llr_sc}
$$
\hat{L}_i \approx \max_{\bar{c}^{(i)} \in  \langle G_\ell^{(i) }\rangle \,:\, \bar{c}^{(i)}_\ell = 0}
\sum_{j=0}^{\ell-1} F(L_j \mid \bar{c}^{(i)}_j )
-
\max_{\bar{c}^{(i)} \in \langle G_\ell^{(i)}\rangle \,:\, \bar{c}^{(i)}_\ell = 1}
\sum_{j=0}^{\ell-1} F(L_j \mid \bar{c}_j^{(i)}).
$$
    
\bibliographystyle{ieeetr}
\bibliography{main_bib}

\end{document}